\documentclass[a4paper,11pt]{article}

\usepackage{jheppub} 

\usepackage[T1]{fontenc} 
\usepackage{amsmath,amssymb,bm}
\usepackage{hyperref}
\usepackage{mathtools}
\usepackage{physics}

\usepackage{array}
\usepackage{tikz}
\usetikzlibrary{calc,positioning,arrows.meta}
\newcommand{\diag}{\operatorname{diag}}

\newcommand{\Mcal}{\mathcal M}

\title{\boldmath Toward a Three-Body Holonomy Standard Model}

\author{Tadashi Yoshikawa,\note{Corresponding author.}}
\affiliation{Nagoya Aoi University \\Nagoya, Japan}
\emailAdd{tyoshika@nagoya-aoi.ac.jp}

\abstract{
We propose a phenomenological framework toward a
Three-Body Holonomy Standard Model in which
the number three is selected by the minimal
primitive cyclic structure of a democratic
many-body interaction network.
The resulting threefold holonomy information
is assumed to survive in $(3+1)$ dimensions
as an internal three-state space.

A Lorentz-invariant six-fermion interaction
provides a candidate microscopic origin of
the internal cyclic correlations, with their
matching to Wilson holonomy left as a
dynamical requirement.
The parity-related holonomy sectors of the
underlying $(1+1)$-dimensional Dirac system
motivate left- and right-handed internal
structures and a trinification-like
$SU(3)_C\times SU(3)_L\times SU(3)_R$
gauge completion.

Vertical holonomy selects a $2+1$ splitting,
while spontaneous left-right vacuum
misalignment provides a possible route to
chiral symmetry breaking.
Horizontal holonomy organizes the three-family
mass structure, with heavy-fermion mixing
generating effective flavor and CP-violating
patterns.

Although a complete ultraviolet derivation
remains open, the framework provides a
unified scenario connecting primitive
three-body dynamics to gauge symmetry,
symmetry breaking, and fermion flavor.
}

\begin{document}
\maketitle
\flushbottom

\section{Introduction}

The Standard Model repeatedly displays the number three. There are three colors,
three fermion generations, and three-state structures in flavor space.
Most model-building approaches begin by postulating a gauge or family
symmetry and then engineering its breaking.  Here we investigate the
opposite logical order.  The central question is whether the number three
can be selected at a more primitive level and subsequently inherited by
the low-energy gauge and flavor structures. It seems that "nature chooses three".

The relativistic three-body Dirac system in
$(1+1)$ dimensions was investigated by
Sakamoto et al.~\cite{Munakata1990,Sakamoto1993,SakamotoNakanoYoshikawa1993}.
Topologically nontrivial three-body contact
interactions were subsequently explored
by Ohya~\cite{Ohya2024}.
More recently, genuine three-body holonomy
was shown to enter the invariant mass spectrum
through phase-sector mixing in a relativistic
Dirac system~\cite{YoshikawaPhase2026}.

In our previous work~\cite{YoshikawaFamily2026},
we proposed that the relative-position information
of three particles can be encoded in a cyclic
holonomy, giving rise to a threefold mass spectrum,
\begin{equation}
 M^{[k]}
 =
 M_0+
 2\Delta
 \cos\left(
 \frac{\theta_3+2\pi k}{3}
 \right),
 \qquad k=0,1,2.
\end{equation}
Within this minimal cyclic construction,
$N=3$ is distinguished as the smallest number
of particles capable of forming a nontrivial
closed loop. Furthermore, the two opposite orientations
of the three-body holonomy are related by parity.
They give rise to parity-related sectors,
which are coupled by the constituent-mass operator
in the $(1+1)$-dimensional Dirac system.
Thus, three essential ingredients emerge. There are 
a distinguished three-body cyclic structure,
a threefold mass hierarchy, and a mass-induced
coupling between parity-related sectors.
The last feature is reminiscent of the coupling
between left- and right-handed fermionic
components through a Dirac mass term.

These observations suggest that an analogous
mechanism may also be realized in $(3+1)$
dimensions. However, the direct extension
of the $(1+1)$-dimensional three-body holonomy
is not straightforward, since its construction
relies on the ordering of particle positions
along a single spatial dimension.

We therefore adopt a different viewpoint.
In $(1+1)$ dimensions, the three-body holonomy
encodes information associated with the relative
positions of three interacting particles.
This suggests that the underlying three-body
interaction, rather than the one-dimensional
particle ordering itself, may provide the
physical origin of the corresponding internal
structure in higher dimensions.

Motivated by this observation, we hypothesize
that genuine three-body interactions in
$(3+1)$ dimensions can generate an effective
three-state internal space, in which the
information inherited from the three-body
interaction is represented by a cyclic holonomy.
This provides the starting point for exploring
the emergence of threefold fermion structures,
mass hierarchies, and left-right sectors
within a unified framework.

We do not assume that the very early theory contains only a three-body
interaction.  At a sufficiently high scale $\Lambda_*$, generic many-body
interactions may coexist,
\begin{equation}
{\cal L}_{\rm UV}
=
{\cal L}_2
+
\sum_{N\ge3}
\frac{G_N}{\Lambda_*^{3N-4}}{\cal O}_{2N},
\qquad
{\cal O}_{2N}\sim(\bar\chi\Gamma\chi)^N,
\label{eq:UVmanybodyIntro}
\end{equation}
where the detailed Lorentz and internal contractions are left open.
Higher-$N$ local operators are increasingly irrelevant in the infrared.
The proposal made here is more specific. Among the many-body sectors, the
three-body sector is the first one that can carry one and only one
independent primitive holonomy on a democratic complete interaction
network.  For $N>3$ several independent cycles occur, and their holonomies
do not in general select a unique common eigenbasis without additional
dynamical information.  Thus the three-body interaction need not remain
large at low energy. Rather, its dimensionless closed-loop information can
survive as an effective holonomy memory.

Our central hypothesis is therefore that,
although generic many-body interactions
may exist at high energies, the minimal
three-body cyclic structure can retain
its holonomy information in the effective
low-energy description.

This hypothesis provides a possible
dynamical origin for a three-state
internal space in $(3+1)$ dimensions,
without postulating an $SU(3)$ gauge
symmetry at the outset.

This statement is a selection scenario.  
Its purpose is to identify the minimal dynamical
structure that must be demonstrated if the repeated low-energy appearance
of three is to have a common origin.

The logical sequence studied in this paper is
\begin{enumerate}
\item UV N-body interactions $\Rightarrow $
  three-body sector as the unique one-cycle many-body sector.
\item Genuine three-body dynamics $\Rightarrow $
  $W_3=e^{i\theta_3} \Rightarrow 
  \mathcal H_{\rm int}\simeq\mathbb C^3 $. \cite{YoshikawaFamily2026,YoshikawaSU32026}
\item Lorentz-invariant six-fermion cyclic channels
  $\Rightarrow$ candidate dynamical origin of internal
  transport links and Wilson holonomy.
\item Three-state carrier + local internal covariance
  $\Rightarrow $ non-Abelian $SU(3)$ completion.
\item parity-paired inherited carriers
  $\Rightarrow $ $SU(3)_L\times SU(3)_R$.
\item $\langle W_C\rangle\propto I_3
  \Rightarrow  SU(3)_C$\ unbroken.
\item $T_8$-type vertical splitting and vacuum selection
  $\Rightarrow $ $SU(2)_L\times U(1)_Y$. \cite{YoshikawaSU32026}
\item Horizontal holonomy + heavy-fermion deformation
  $\Rightarrow $ $M_f^{\rm eff},\ V_{\rm CKM},\ U_{\rm PMNS}$.
\end{enumerate}
The resulting quark and lepton mixing structures
are described by the CKM and PMNS matrices,
respectively, whose theoretical foundations
were established in
Refs.~\cite{KobayashiMaskawa1973,MNS1962}.

Holonomy as a physical carrier of closed-path information is well
established in geometric-phase and gauge-theory settings
\cite{Simon1983,Berry1984,WilczekZee1984,Hosotani1983,Hosotani1984}.
The appearance of internal gauge structure from parallel transport in a
degenerate state space is particularly relevant to the present logic
\cite{WilczekZee1984}, while three-body systems provide examples in which
geometric phases and internal dynamics are intertwined \cite{Lin1997}.
There is also a useful analogy with few-body effective field theory in
the universal large-scattering-length problem, genuinely new information
first appears in the three-body sector, while parts of higher-body physics
are tied to the three-body scale \cite{Bedaque1999,BraatenHammer2006}.
We do not identify that mechanism with the present holonomy construction.
It is cited only as a physical precedent for irreducible three-body
information becoming the organizing datum of higher-body observables.

Building on these developments, we investigate
whether the resulting three-body cyclic
information can be inherited by a finite
three-state internal structure in an effective
$(3+1)$-dimensional description.
This promotion is an inheritance hypothesis
rather than a complete microscopic derivation.

A central issue in this dimensional extension is how
genuine three-body information can be represented by
Lorentz-invariant interactions and subsequently encoded
in an internal holonomy.  We examine a local six-fermion
operator constructed from three Dirac bilinears,
whose cyclic contractions contain the two oppositely
oriented three-site channels.  In the presence of suitable
bilinear correlations, this interaction contributes to
an effective three-state mass structure.  If the resulting
channel amplitudes admit a unitary parallel-transport
description, their closed product defines an internal
Wilson holonomy\cite{Wilson1974, WilczekZee1984} 
with gauge-invariant spectral information.

The dynamical matching between the six-fermion
interaction and the effective internal transport
links remains an open problem.

\section{Overall framework: from UV many-body dynamics to low-energy flavor}
\label{sec:overview}

This section gives the map of the construction before entering the detailed
calculations.  The key point is that the number three is not inserted at a
single late stage to fit three generations. It is required to propagate
through the entire chain.

\subsection{UV many-body interactions and the surviving holonomy memory}

At the highest scale considered here, we allow generic $N$-body operators
as in Eq.~\eqref{eq:UVmanybodyIntro}.  A local fermionic $N$-body operator
has canonical dimension $3N$ in four dimensions and therefore carries a
coefficient suppressed by $\Lambda_*^{-(3N-4)}$.  For comparable
dimensionless Wilson coefficients, direct effects from $N>3$ operators
are consequently more strongly suppressed as the system flows to lower
energies.  This by itself does not privilege the three-body sector, since
the six-fermion operator is also irrelevant.  The additional ingredient is
that the three-body sector can leave a dimensionless closed-loop datum,
\begin{equation}
W_3=U_{12}U_{23}U_{31}=e^{i\theta_3},
\label{eq:W3overview}
\end{equation}
even after the local interaction that generated it has become weak in the
low-energy description.  In this sense the interaction amplitude may
decouple while its holonomy information is retained.

The selection of three is then tied to the number of independent cycles.
For a democratic complete graph $K_N$,
\begin{equation}
\beta_1(K_N)=\frac{(N-1)(N-2)}{2}.
\label{eq:betaOverview}
\end{equation}
Thus $N=2$ carries no independent primitive cycle, $N=3$ carries exactly
one, and $N>3$ carries several.  If the low-energy state is required to be
organized by a single primitive holonomy eigenbasis, $N=3$ is the unique
minimal possibility.  For $N>3$, noncommuting holonomies cannot in general
be simultaneously diagonalized, while even mutually commuting holonomies
require several independent phases.  Higher-body interactions can therefore
exist in the UV without furnishing an additional uniquely selected
low-energy eigenbasis.

The proposal is summarized by
\begin{align}
\text{UV many-body interactions}
& \;\xrightarrow[\text{RG / decoupling}]{}\; 
\text{three-body holonomy memory} \nonumber \\
& \;\xrightarrow{}\;
\text{one primitive three-state eigenbasis}.
\label{eq:UVto3}
\end{align}
We emphasize that Eq.~\eqref{eq:UVto3} is a model hypothesis to be tested
by an explicit ultraviolet completion, not a claim that all possible
higher-body holonomies vanish mathematically.

\subsection{From the three-state carrier to gauge and flavor structures}

The finite holonomy spectrum is promoted to
$\mathcal H_{\rm int}\simeq\mathbb C^3$.  The two opposite branches of the
relativistic three-body problem motivate the inherited pair
$\mathbb C_L^3\oplus\mathbb C_R^3$.  When local changes of basis in these
carriers are treated as redundancies, gauge connections are required and
their minimal non-Abelian parts are $SU(3)_L$ and $SU(3)_R$.  Color is an
independent locally gauged three-state carrier, $SU(3)_C$, whose holonomy is
assumed proportional to the identity so that color remains unbroken.

The horizontal three-state carrier is different.  Its cyclic operator
$C_H$ defines a physical family eigenbasis and is therefore retained as a
spectral holonomy rather than promoted to an additional local family gauge
group.  This distinction is important.

\subsection{Mass matrices and the fitting map}

The horizontal holonomy first generates a restricted zeroth-order mass
matrix rather than a general complex $3\times3$ texture.  For each light
fermion sector $f$ we define
\begin{equation}
M_f^{(0)}=m_{0f}
\left[I+r_f(C_H+C_H^\dagger)+s_f(C_H^2+C_H^{\dagger2})\right],
\label{eq:M0overview}
\end{equation}
with $C_H^3=e^{i\theta_H}I$.  The parameters
\begin{equation}
\{m_{0f},r_f,s_f,\theta_H\}
\end{equation}
therefore belong to the holonomy backbone.  
The threefold cyclic holonomy also provides a
natural mass-splitting structure. Its three
eigenvalue branches take the form
\begin{equation}
M^{[k]}=M_{0f}+2\Delta_f
\cos\left(\frac{\theta+2\pi k}{3}\right)+2\Delta^{(2)}_f \cos\left(\frac{2\theta+4\pi k}{3}\right),
\qquad k=0,1,2.
\label{eq:intro_mass_splitting}
\end{equation}
At $\theta=0$, the spectrum becomes
$(M_0+2\Delta +2s,M_0-\Delta-s,M_0-\Delta-s)$,
exhibiting a characteristic $2+1$ structure.
After a suitable permutation of states, the
traceless splitting is proportional to
$T_8=\operatorname{diag}(1,1,-2)/(2\sqrt3)$.
This spectral structure motivates the
vertical symmetry-breaking direction discussed
below, while the horizontal holonomy organizes
the three-family mass hierarchy.
Heavy vertical states then
deform this matrix.  In the down sector, for example,
\begin{equation}
{\cal M}_{dD}=
\begin{pmatrix}
M_d^{(0)} & X_D\\
Y_D & M_D
\end{pmatrix},
\qquad
M_d^{\rm eff}=M_d^{(0)}-X_DM_D^{-1}Y_D+\cdots.
\label{eq:MdFitOverview}
\end{equation}
The numerical fit is performed on the effective light matrices
$M_u^{\rm eff}$ and $M_d^{\rm eff}$, not directly on $H_H$.

Writing
\begin{equation}
{\cal H}_q=M_q^{\rm eff}M_q^{{\rm eff}\dagger},
\qquad
U_{qL}^\dagger {\cal H}_q U_{qL}
=\diag(m_{q1}^2,m_{q2}^2,m_{q3}^2),
\end{equation}
the physical quark mixing matrix is
\begin{equation}
V_{\rm CKM}=U_{uL}^\dagger U_{dL}.
\label{eq:CKMoverview}
\end{equation}
If both sectors contained only the common circulant holonomy backbone,
they would share the same Fourier eigenvectors and
$V_{\rm CKM}^{(0)}=I$.  The observed mixing therefore measures the
misalignment generated by sector-dependent deformations, especially the
heavy-fermion terms.  Complex relative phases in those deformations are
the immediate source of the physical CKM phase. In a more predictive UV
completion they should themselves be inherited from the primitive holonomy
rather than introduced independently.

For a small deformation, the propagation from a matrix parameter to a CKM
element can be displayed analytically.  In the common holonomy eigenbasis,
\begin{equation}
(\Omega_q)_{ij}
\simeq
\frac{(\widetilde{\delta{\cal H}}_q)_{ij}}
{m_{qj}^{(0)2}-m_{qi}^{(0)2}},
\qquad i\ne j,
\end{equation}
and hence
\begin{equation}
V_{\rm CKM}\simeq I+\Omega_d-\Omega_u.
\label{eq:CKMpropOverview}
\end{equation}
This relation makes explicit how the heavy-sector parameters propagate into
$V_{us}$, $V_{cb}$, $V_{ub}$, and the rephasing-invariant CP measure.
The same strategy is used in the lepton sector.

The complete conceptual map is therefore
\begin{equation}
\begin{array}{c}
\text{UV }N\text{-body dynamics}\\
\downarrow\\
\text{RG suppression + one-cycle holonomy selection}\\
\downarrow\\
\text{genuine three-body holonomy }W_3\\
\downarrow\\
\mathbb C_L^3\oplus\mathbb C_R^3\ \text{and gauge completion}\\
\downarrow\\
\text{vertical }2+1\ \text{breaking + horizontal family holonomy}\\
\downarrow\\
M_f^{(0)}+\delta M_f^{\rm heavy}\\
\downarrow\\
\text{masses, CKM, PMNS, and CP phases}.
\end{array}
\label{eq:overallmap}
\end{equation}

\section{Why three? A unique one-cycle complete network}
\label{sec:whythree}

\subsection{The complete pairwise network and its cycle rank}

To compare different body numbers without imposing a preferred ring by
hand, we begin with the most symmetric graph-theoretic idealization. Every
pair of constituents is allowed to communicate.  The underlying network
is then the complete graph $K_N$, with
\begin{equation}
V=N,
\qquad
E=\frac{N(N-1)}{2}.
\end{equation}
For a connected graph the number of independent cycles is the cycle rank,
or first Betti number
\begin{equation}
\beta_1=E-V+1.
\end{equation}
where (V) and (E) denote, respectively, the numbers of vertices and edges of the corresponding graph.
Hence
\begin{equation}
\beta_1(K_N)=\frac{(N-1)(N-2)}{2}.
\label{eq:bettiKN}
\end{equation}
This gives
\begin{equation}
\beta_1(K_1)=0,
\qquad
\beta_1(K_2)=0,
\qquad
\beta_1(K_3)=1,
\qquad
\beta_1(K_4)=3.
\label{eq:betti1234}
\end{equation}
Thus the distinction between three and four is not merely that both admit
closed paths.  In the complete interaction network, three has exactly one
independent loop, whereas four already requires three independent loop
data.

The use of $K_N$ is itself a physical minimality assumption before a
symmetry-breaking direction is selected, no pairwise link is privileged.
If a microscopic theory removes links or enforces a special graph, a
different cycle rank can of course result.  Our claim concerns the
unbiased complete pairwise network.

\subsection{$N=1$ and $N=2$: no independent primitive holonomy}

For $N=1$ there is no inter-site transport and therefore no nontrivial
closed cycle.  For $N=2$ the only elementary motion is a back-and-forth
path,
\begin{equation}
1\to2\to1,
\end{equation}
and the pure exchange operator satisfies
\begin{equation}
P_2=P_2^{-1}.
\end{equation}
The two-site system can possess symmetric and antisymmetric eigenmodes,
but it does not contain an independent oriented primitive loop.  This is
consistent with $\beta_1(K_1)=\beta_1(K_2)=0$.

\subsection{$N=3$: one loop, two orientations, one threefold eigenbasis}

For three constituents the complete graph $K_3$ is itself a triangle and
contains exactly one independent cycle,
\begin{equation}
\gamma_3:
1\to2\to3\to1.
\end{equation}
Its inverse,
\begin{equation}
\gamma_3^{-1}:
1\to3\to2\to1,
\end{equation}
is not a second independent cycle. It is the opposite orientation of the
same primitive loop.  Thus
\begin{equation}
N=3:
\qquad
\text{one independent loop}
+
\text{two orientations}.
\label{eq:oneLoopTwoOrientations}
\end{equation}

Introduce the cyclic transport operator
\begin{equation}
C_3=
\begin{pmatrix}
0&0&e^{i\theta_3}\\
1&0&0\\
0&1&0
\end{pmatrix},
\qquad
C_3^3=e^{i\theta_3}I.
\label{eq:C3}
\end{equation}
Its eigenvalues are
\begin{equation}
\lambda_k=
\exp\left[i\frac{\theta_3+2\pi k}{3}\right],
\qquad k=0,1,2,
\label{eq:lambda3}
\end{equation}
with eigenstates
\begin{equation}
|k;3\rangle
=
\frac{1}{\sqrt3}
\begin{pmatrix}
1\\ e^{-iq_k}\\ e^{-2iq_k}
\end{pmatrix},
\qquad
q_k=\frac{\theta_3+2\pi k}{3}.
\label{eq:states3}
\end{equation}
The inverse orientation maps the eigenphases to the opposite orientation,
up to relabeling of the three branches.  A single primitive holonomy and
its inverse therefore organize the entire minimal three-state carrier.
This is the sense in which the three-body system is spectrally
self-contained.

The threefold eigenphases lead naturally to a
mass-splitting structure. Consider the minimal
Hermitian cyclic operator
\begin{equation}
{\cal M}_3
=
M_0 I+\Delta(C_3+C_3^\dagger),
\label{eq:primitiveMassOperator}
\end{equation}
where $M_0$ and $\Delta$ are real parameters.
Since $C_3|k;3\rangle=e^{iq_k}|k;3\rangle$,
the three eigenvalues are
\begin{equation}
M^{[k]}
=
M_0+2\Delta
\cos\left(\frac{\theta_3+2\pi k}{3}\right),
\qquad k=0,1,2.
\label{eq:primitiveMassSplitting}
\end{equation}
Thus a single cyclic holonomy phase generates
three distinct spectral branches.

In the special limit $\theta_3=0$, the
spectrum becomes
\begin{equation}
(M^{[0]},M^{[1]},M^{[2]})
=
(M_0+2\Delta,M_0-\Delta,M_0-\Delta).
\label{eq:primitiveTwoPlusOne}
\end{equation}
The twofold degeneracy exhibits a natural
$2+1$ spectral structure.
Its traceless part can be represented by
a $T_8$-type diagonal matrix after a
permutation of the eigenstates.

For nonzero $\theta_3$, the degeneracy is
generically lifted. The physical masses
associated with a Dirac mass operator are
the corresponding nonnegative singular values.
The dynamical interpretation of this splitting
in the vertical and horizontal sectors is
discussed below.

\subsection{$N=4$: several independent holonomies are already required}

For the complete four-body network $K_4$,
\begin{equation}
\beta_1(K_4)=3.
\end{equation}
A convenient cycle basis may be chosen, for example, as
\begin{equation}
\gamma_{123}=1\to2\to3\to1,
\qquad
\gamma_{124}=1\to2\to4\to1,
\qquad
\gamma_{134}=1\to3\to4\to1.
\label{eq:K4cycles}
\end{equation}
The associated holonomies are
\begin{equation}
W_{123},
\qquad
W_{124},
\qquad
W_{134}.
\end{equation}
In an Abelianized description this already requires three independent
loop phases, modulo dynamical relations.  In a non-Abelian description
the holonomies need not commute,
\begin{equation}
[W_{\gamma_a},W_{\gamma_b}]\neq0,
\end{equation}
so a common eigenbasis need not exist.  Therefore a generic complete
four-body network is not organized by one primitive holonomy operator.
Additional loop information is unavoidable.

The contrast is therefore
\begin{equation}
\beta_1=
\begin{cases}
0, & N=1,2,\\
1, & N=3,\\
>1, & N>3
\end{cases}
\quad\text{for }K_N.
\label{eq:bettiSelection}
\end{equation}
Among complete pairwise networks, three is the unique body number with
exactly one independent closed cycle.  This gives a sharper version of
the statement ``three is the smallest oriented loop''. Three is also the
only minimal complete network for which one primitive holonomy can carry
the full loop information.

\subsection{Finite site cost as a secondary dynamical selection}

The topological argument already distinguishes $N=3$ from both smaller
and larger complete networks.  A finite site cost can reinforce the same
selection dynamically.  We retain the effective ansatz
\begin{equation}
E_N=
\mu N
-A M\left(\frac{t}{M}\right)^N,
\qquad N\ge3,
\qquad 0<\frac{t}{M}<1.
\label{eq:EN}
\end{equation}
Then
\begin{equation}
E_{N+1}-E_N
=
\mu
+A M\left(\frac{t}{M}\right)^N
\left(1-\frac{t}{M}\right)>0,
\end{equation}
so that
\begin{equation}
E_3<E_4<E_5<\cdots.
\end{equation}
The logical roles of the two ingredients should be kept distinct:
\begin{equation}
\begin{aligned}
&\text{cycle topology of }K_N:
&&N=3\text{ is the unique one-cycle complete network},\\
&\text{finite site cost}:
&&\text{larger carriers are energetically disfavored}.
\end{aligned}
\label{eq:why3}
\end{equation}
Equation~(\ref{eq:EN}) remains an effective ansatz whose microscopic
origin must ultimately be derived.

\section{From one-dimensional holonomy to internal $L/R$ carriers}
\label{sec:LRorigin}

\subsection{Why the $(1+1)$-dimensional realization cannot be lifted literally}

The original three-body holonomy is realized in a one-dimensional
configuration space.  For distinguishable particles with coincidence
points removed, one spatial dimension is special because particle order
cannot be changed continuously without crossing a coincidence surface.
The collision-free configuration space therefore separates into ordering
sectors such as
\begin{equation}
x_1<x_2<x_3,
\qquad
x_1<x_3<x_2,
\qquad\ldots
\label{eq:orderingSectors}
\end{equation}
The low-dimensional contact problem can consequently encode physical
information in phase relations between such sectors.

In three spatial dimensions this geometric mechanism does not survive in
the same form.  Particles can move around one another without passing
through a coincidence point, so the one-dimensional ordering sectors are
not disconnected spatial sectors of the $(3+1)$-dimensional theory.  The
configuration-space realization of the holonomy therefore cannot simply
be reinterpreted as an ordinary spatial loop in four dimensions.

This observation creates a concrete requirement.  If the finite spectral
information generated by the primitive three-body holonomy is to survive
when the one-dimensional ordering geometry is lost, that information
must be carried by a different degree of freedom.  We take the minimal
possibility to be an internal carrier:
\begin{equation}
\begin{aligned}
&\text{one-dimensional configuration-space holonomy}\\
&\hspace{18mm}\longrightarrow
\text{ finite internal holonomy data in }(3+1)\text{ dimensions}.
\end{aligned}
\label{eq:internalNecessity}
\end{equation}
Thus the geometry itself is not inherited, The finite holonomy spectrum
is.

\subsection{The inherited three-state carrier}

The primitive three-cycle has the three spectral branches
$|k;3\rangle$, $k=0,1,2$, of Eq.~(\ref{eq:states3}).  Once the spatial
ordering interpretation is no longer available, the minimal finite
Hilbert space capable of retaining these three orthogonal labels is
\begin{equation}
\mathcal H_{\rm hol}
=\operatorname{span}\{|0\rangle,|1\rangle,|2\rangle\}
\simeq\mathbb C^3.
\label{eq:C3carrier}
\end{equation}
We therefore identify the effective $(3+1)$-dimensional carrier as
\begin{equation}
\mathcal H_{\rm int}\simeq\mathbb C^3.
\label{eq:HintC3}
\end{equation}
This is not a derivation of an $SU(3)$ gauge theory from the
one-dimensional model.  It is the minimal inheritance rule that preserves
the threefold holonomy spectrum after its original spatial carrier has
ceased to exist.

Basis changes on the carrier act as
\begin{equation}
\Psi(x)\to U(x)\Psi(x),
\qquad U(x)\in U(3).
\end{equation}
If this basis covariance is promoted to a local redundancy in the
effective theory, its non-Abelian part is $SU(3)$.  The gauge
interpretation is therefore a further effective-theory step, logically
after the three-state carrier itself.

\subsection{Two massless sectors in the relativistic three-body system}

In the relativistic $(1+1)$-dimensional three-body construction, define
\begin{equation}
Q_3=\alpha_1\alpha_2\alpha_3,
\qquad Q_3^2=1.
\label{eq:Q3}
\end{equation}
The kinetic and pair-interaction parts commute with $Q_3$ in the
reference system, so that in the massless limit
\begin{equation}
[H_0,Q_3]=0.
\end{equation}
The Hilbert space separates as
\begin{equation}
\mathcal H=\mathcal H_+\oplus\mathcal H_-,
\qquad
Q_3|\Psi_\pm\rangle=\pm|\Psi_\pm\rangle.
\label{eq:Hplusminus}
\end{equation}
A genuine three-body holonomy generated by $Q_3$ may be written as
\begin{equation}
W_3=e^{i\theta_3Q_3},
\end{equation}
so that
\begin{equation}
W_3|\Psi_\pm\rangle=e^{\pm i\theta_3}|\Psi_\pm\rangle.
\label{eq:oppositephases}
\end{equation}
The massless theory therefore contains two dynamically decoupled sectors
carrying opposite holonomy phases.

\subsection{Parity exchanges the two holonomy branches}

Under spatial parity in one dimension,
\begin{equation}
x_i\to -x_i,
\qquad
P\alpha_iP^{-1}=-\alpha_i.
\end{equation}
Because $Q_3$ contains an odd number of velocity operators,
\begin{equation}
P Q_3P^{-1}
=(-\alpha_1)(-\alpha_2)(-\alpha_3)
=-Q_3.
\label{eq:PQ3}
\end{equation}
Hence parity exchanges the two sectors,
\begin{equation}
P:\quad \mathcal H_+\longleftrightarrow\mathcal H_-.
\label{eq:Pexchange}
\end{equation}
Equivalently,
\begin{equation}
P W_3P^{-1}=W_3^{-1},
\end{equation}
which maps
\begin{equation}
e^{+i\theta_3}\longleftrightarrow e^{-i\theta_3}.
\end{equation}
Thus the two primitive holonomy orientations form a parity-paired
structure before any four-dimensional chiral interpretation is imposed.

\subsection{The constituent mass connects the two modes}

The constituent-mass operator anticommutes with $Q_3$,
\begin{equation}
\{H_m,Q_3\}=0.
\label{eq:massanticommute}
\end{equation}
Therefore
\begin{equation}
H_m:\mathcal H_+\to\mathcal H_-,
\qquad
H_m:\mathcal H_-\to\mathcal H_+.
\end{equation}
In the $Q_3$ basis, the Hamiltonian has the schematic form
\begin{equation}
H=
\begin{pmatrix}
H_+&\Mcal\\
\Mcal^\dagger&H_-
\end{pmatrix},
\label{eq:blockH}
\end{equation}
where the off-diagonal block vanishes in the massless limit.  The physical
picture is therefore
\begin{equation}
\begin{aligned}
&\text{massless: }e^{+i\theta_3}\text{ and }e^{-i\theta_3}
\text{ are separated modes},\\
&\text{massive: the mass operator connects the two modes}.
\end{aligned}
\label{eq:massconnects}
\end{equation}
This is the low-dimensional precursor of a left--right mass coupling.

\subsection{Inheritance as left- and right-handed internal sectors}

We now combine the two pieces of inherited information: each primitive
holonomy branch carries the threefold internal spectrum, and parity
exchanges the two branches.  The minimal $(3+1)$-dimensional carrier is
therefore taken to be
\begin{equation}
\mathcal H_+\leadsto\mathcal H_L\simeq\mathbb C_L^3,
\qquad
\mathcal H_-\leadsto\mathcal H_R\simeq\mathbb C_R^3,
\label{eq:inheritLR}
\end{equation}
with
\begin{equation}
\mathcal H_{\rm int}
=\mathbb C_L^3\oplus\mathbb C_R^3.
\label{eq:HintLR}
\end{equation}
At the level of the inherited holonomy operators we therefore write
\begin{equation}
W_L\equiv W_3,\qquad
W_R\equiv P W_L P^{-1}=W_L^{-1}=W_L^\dagger,
\label{eq:WLRparity}
\end{equation}
where the last equality assumes unitary parallel transport.  Equation~\eqref{eq:WLRparity}
is the precise sense in which the two four-dimensional internal branches
inherit the parity pairing of the original three-body system.

This is \emph{not} a literal identification of the $(1+1)$-dimensional
operator $Q_3$ with the four-dimensional chirality operator $\gamma_5$.
Rather, it is an inheritance of three structural facts:
\begin{enumerate}
\item the primitive holonomy has two opposite orientations;
\item parity exchanges the two orientations;
\item a mass operator connects the two otherwise separated sectors.
\end{enumerate}

Local basis transformations act as
\begin{equation}
\Psi_L\to U_L\Psi_L,
\qquad
\Psi_R\to U_R\Psi_R,
\end{equation}
with $U_{L,R}\in U(3)_{L,R}$.  If these are promoted to local internal
redundancies, the corresponding non-Abelian structure is
\begin{equation}
SU(3)_L\times SU(3)_R.
\label{eq:SU3LR}
\end{equation}
The ordinary Dirac-type mass coupling then has the expected form
\begin{equation}
\mathcal L_m
=-\overline\Psi_L M\Psi_R
-\overline\Psi_R M^\dagger\Psi_L.
\label{eq:LRmass}
\end{equation}
The route from the low-dimensional problem to the effective chiral
carrier is therefore
\begin{equation}
\begin{aligned}
&(1+1)\text{D ordering sectors}
\longrightarrow
\text{three-body holonomy spectrum}
\\
&\longrightarrow
\mathbb C^3\text{ internal carrier}
\longrightarrow
\mathbb C_L^3\oplus\mathbb C_R^3
\\
&\longrightarrow
SU(3)_L\times SU(3)_R
\quad\text{(upon local gauging).}
\end{aligned}
\label{eq:dimensionalBridge}
\end{equation}
The crucial point is that the first arrow involving $(3+1)$ dimensions is
an explicitly stated inheritance hypothesis: the spatial ordering
geometry is discarded, while its finite holonomy data are retained.

\section{From three-body holonomy to gauge connection and natural $SU(3)$ structure}
\label{sec:gaugeconnection}

The previous sections treated the holonomy as the primitive carrier of
closed three-body information and then retained its finite spectrum as an
internal degree of freedom in $(3+1)$ dimensions.  We now make explicit the
next logical step.  A holonomy is not an isolated phase assigned to a loop.
It is the integrated record of a rule for parallel transport.  Once the
inherited carrier is allowed to have a space-time dependent local basis,
that rule is represented by a connection.  This is the precise sense in
which gauge structure enters the present construction.  Geometric and
non-Abelian holonomies in quantum systems provide the standard background
for this relation~\cite{Berry1984,Simon1983,WilczekZee1984}, while Wilson-line
vacua illustrate how a connection holonomy can select a gauge-symmetry
breaking direction~\cite{Hosotani1983,Hosotani1984}.

\subsection{Six-fermion cyclic channels as a candidate microscopic origin}
\label{subsec:sixpoint-origin}

The ordering information that underlies the $(1+1)$-dimensional construction
cannot be interpreted literally as a left-to-right spatial ordering in
$(3+1)$ dimensions.  We therefore distinguish \emph{spatial ordering} from
\emph{cyclic ordering of internal interaction channels}.  The latter admits
a Lorentz-invariant six-fermion realization and provides a concrete
candidate origin for the three links used below\cite{tHooft1976, Klevansky1992}.  
This subsection is a
matching proposal, not a derivation of gauge dynamics from a local vertex.

For three Dirac flavors $\chi_i$ define the Lorentz-scalar bilinears
\begin{equation}
 B_{ij}=\bar\chi_{Ri}\chi_{Lj},\qquad
 {\cal L}_{6}=\frac{\kappa_6}{\Lambda_6^5}\det B
 +\frac{\kappa_6^*}{\Lambda_6^5}(\det B)^\dagger.
 \label{eq:sixpointdet}
\end{equation}
Since each $B_{ij}$ contains two fermionic fields, $\det B$ is a
local six-fermion operator of canonical dimension nine.  Its precise
gauge embedding is model dependent: the fields and their contractions
must be chosen so that Eq.~\eqref{eq:sixpointdet} is invariant under
the ultraviolet gauge group.  In particular the hidden fermions here
are not automatically identical to the Standard-Model quarks or to the
heavy $D,N$ partners introduced later.

The determinant contains all six permutations,
\begin{align}
 \det B={}&B_{11}B_{22}B_{33}
 +B_{12}B_{23}B_{31}+B_{13}B_{21}B_{32}\nonumber\\
 &-B_{11}B_{23}B_{32}-B_{12}B_{21}B_{33}
 -B_{13}B_{22}B_{31}.
 \label{eq:sixpointpermutations}
\end{align}
Of particular interest are the two opposite three-cycles,
\begin{equation}
 {\cal O}_{123}=B_{12}B_{23}B_{31},\qquad
 {\cal O}_{132}=B_{13}B_{32}B_{21}.
 \label{eq:sixpointcycles}
\end{equation}
They are distinguished by \emph{internal index routing}, not by a
preferred spatial direction.  The two three-cycles are both even
permutations and are not, in general, Hermitian conjugates.  Consequently,
one must not equate them directly with the $Q_3=\pm1$ sectors of the
$(1+1)$-dimensional Dirac Hamiltonian.  The construction only identifies
a possible internal channel structure to which the earlier holonomy
information could be matched.

If a matrix of bilinear correlations develops,
$\Sigma_{ij}=\langle B_{ij}\rangle$, expansion around this background
gives a bilinear term proportional to the cofactor of $\Sigma$:
\begin{equation}
 \delta {\cal L}_{6}
 =\frac{\kappa_6}{\Lambda_6^5}
 \sum_{ij}{\rm Cof}_{ij}(\Sigma)\,\delta B_{ij}
 +\mathrm{h.c.}+O((\delta B)^2),
 \label{eq:sixpointcofactor}
\end{equation}
where $\delta B=B-\Sigma$.  For instance,
\begin{equation}
 {\rm Cof}_{12}(\Sigma)
 =\Sigma_{23}\Sigma_{31}-\Sigma_{21}\Sigma_{33}.
 \label{eq:sixpointcofactor12}
\end{equation}
Thus a six-point interaction can feed its cyclic correlations into a
three-by-three \emph{Dirac} mass matrix.  This mean-field expansion
neither proves the existence of the condensate nor makes its phases
basis invariant under independent $U(3)_L\times U(3)_R$ rotations.

The actual internal Wilson-loop construction requires a further,
explicitly specified matching step.  Suppose that the low-energy
single-carrier problem has oriented parallel-transport links
$U_{ij}$, transforming under local frame changes as
\begin{equation}
 U_{ij}\longmapsto G_iU_{ij}G_j^{-1},\qquad
 W_{123}=U_{12}U_{23}U_{31}
 \longmapsto G_1W_{123}G_1^{-1}.
 \label{eq:sixpointmatchlinks}
\end{equation}
Then $\Tr W_{123}$ and the eigenvalues of $W_{123}$ are gauge
invariant, while reversal of the same unitary links gives
$W_{132}=W_{123}^{-1}=W_{123}^{\dagger}$ up to choice of base point.
For scalar unitary links, $W_{123}=e^{i\theta_3}$ and the inverse
orientation has phase $e^{-i\theta_3}$.  The existence of the loop
permits such an invariant phase. It does not compel
$\theta_3\ne0$.  More generally, for fixed Hermitian single-carrier
hoppings $T_{ji}=T_{ij}^*$, the quantity
\begin{equation}
 {J}_{\rm loop}=\operatorname{Im}
 \left(T_{12}T_{23}T_{31}\right)
 \label{eq:sixpointJloop}
\end{equation}
is invariant under three independent site rephasings.  It is a test
for a complex oriented loop in that specified effective basis. Its
nonzero value and its relation to physical weak-basis CP invariants
must still follow from dynamics and matching.  In particular, a
single general Dirac mass matrix has independent left and right basis
rotations, so its naive product of three entries is not itself a
physical CP invariant.

The $(1+1)$-dimensional reference model obeys
\begin{equation}
 [H_0,Q_3]=0,\qquad \{H_m,Q_3\}=0,
 \label{eq:sixpoint1dgrading}
\end{equation}
so that the constituent masses connect the two massless sectors.
In $(3+1)$ dimensions the free Dirac Hamiltonian similarly satisfies
$[\boldsymbol\alpha\cdot\mathbf p,\gamma^5]=0$ and
$\{\beta m,\gamma^5\}=0$.  Nevertheless chirality, orientation of an
internal cycle, and parity of the original configuration sectors are
\emph{different notions}.  Their identification in the full
four-dimensional model is a physical inheritance or matching condition,
not a consequence of Eq.~\eqref{eq:sixpointdet}.
For example, with $\Gamma=(-1)^{N_L}$, one finds
$\Gamma B_{ij}\Gamma^{-1}=-B_{ij}$, and therefore the determinant
six-fermion Hamiltonian is $\Gamma$-odd, just as a Dirac mass
insertion is. It is not automatically the $\Gamma$-even holonomy
operator required for a literal transplantation of the
$(1+1)$-dimensional grading.  Even-order virtual processes may
contribute to a $\Gamma$-even effective Hamiltonian, but their
closed-loop phase must be checked rather than assumed.

Our specific proposal is thus the conditional matching chain
\begin{equation}
 \underbrace{{\cal O}_6}_{\text{six-point vertex}}
 \ \longrightarrow\ 
 \underbrace{\Sigma_{ij},\; T_{ij}}_{\text{dynamical channel matching}}
 \ \longrightarrow\ 
 \underbrace{W_{123}}_{\text{internal parallel transport}}
 \ \longrightarrow\ 
 \underbrace{C_H,\; M_f^{\rm eff}}_{\text{spectral and flavor sector}}.
 \label{eq:sixpointmatchingchain}
\end{equation}
The last arrow is implemented in the existing horizontal-holonomy and
heavy-fermion sectors of this paper. It is not a replacement for the
Wilson-loop gauge construction, nor for the heavy-induced CKM and PMNS
mechanisms.  Demonstrating the first two arrows from a definite
ultraviolet action---including a nontrivial vacuum phase and a
consistent gauge representation---remains an open dynamical task.

\subsection{Discrete parallel transport on the primitive three-cycle}

On the elementary oriented cycle
\begin{equation}
1\longrightarrow2\longrightarrow3\longrightarrow1,
\end{equation}
introduce link variables
\begin{equation}
U_{12},\qquad U_{23},\qquad U_{31}.
\end{equation}
Their ordered product is the primitive closed transport,
\begin{equation}
W_3=U_{12}U_{23}U_{31}.
\label{eq:gaugeW3links}
\end{equation}
Under independent changes of internal basis $G_i$ at the three sites,
\begin{equation}
U_{ij}\longrightarrow G_iU_{ij}G_j^{-1},
\label{eq:discretelinktransform}
\end{equation}
and hence
\begin{equation}
W_3\longrightarrow G_1W_3G_1^{-1}.
\label{eq:W3conjugationnew}
\end{equation}
Therefore the conjugacy class and eigenvalues of $W_3$ are basis
independent.  Equation~\eqref{eq:discretelinktransform} is the discrete
analogue of a gauge transformation.  The logical chain is
\begin{equation}
\text{nontrivial holonomy}
\Longrightarrow
\text{parallel transport law}
\Longrightarrow
\text{connection data}.
\label{eq:holonomyconnectionlogicnew}
\end{equation}
A single loop does not reconstruct a unique microscopic gauge field, but it
already requires a consistent local transport structure.

\subsection{Continuum connection and Wilson holonomy}

In the effective continuum description the links may be represented as
\begin{equation}
U_{ij}=
{\cal P}\exp\!\left(
ig\int_i^j A_\mu dx^\mu
\right),
\end{equation}
so that a closed path $\gamma$ carries the Wilson holonomy
\begin{equation}
W[\gamma]
=
{\cal P}\exp\!\left(
ig\oint_\gamma A_\mu dx^\mu
\right).
\label{eq:Wilsonholonomynew}
\end{equation}
For a local basis transformation $G(x)$,
\begin{equation}
A_\mu\longrightarrow
GA_\mu G^{-1}
-\frac{i}{g}(\partial_\mu G)G^{-1},
\label{eq:connectiontransformationnew}
\end{equation}
and
\begin{equation}
F_{\mu\nu}
=
\partial_\mu A_\nu-\partial_\nu A_\mu
-ig[A_\mu,A_\nu].
\label{eq:curvatureNew}
\end{equation}
For an infinitesimal loop,
\begin{equation}
W[\square]
=I+igF_{\mu\nu}\,\delta S^{\mu\nu}+\cdots,
\end{equation}
showing explicitly that the holonomy contains integrated curvature
information.

This continuum connection should not be identified literally with the
original one-dimensional configuration-space geometry.  The geometric
ordering sectors are lost when the theory is promoted to $(3+1)$
dimensions. What is inherited is the finite transport/spectral information.
The connection in Eq.~\eqref{eq:Wilsonholonomynew} is therefore the local
effective structure that realizes this inherited internal transport.

\subsection{Why $SU(3)$ is natural once $N=3$ is selected}
\label{subsec:naturalSU3}

The number three first selects the carrier dimension,
\begin{equation}
N=3
\quad\Longrightarrow\quad
\mathcal H_3\simeq\mathbb C^3.
\end{equation}
If orthonormal bases of this carrier may be chosen independently at each
space-time point without changing the physics, the natural local basis
redundancy is
\begin{equation}
\Psi(x)\longrightarrow G(x)\Psi(x),
\qquad G(x)\in U(3).
\end{equation}
Accordingly,
\begin{equation}
u(3)=u(1)\oplus su(3),
\end{equation}
and the connection may be decomposed as
\begin{equation}
A_\mu=A_\mu^{(0)}I+A_\mu^aT^a,
\qquad a=1,\ldots,8.
\end{equation}
The natural minimal non-Abelian local structure acting irreducibly on the
three-state carrier is therefore $SU(3)$.  The claim is deliberately
ordered as
\begin{equation}
N=3
\Rightarrow
\mathbb C^3
\Rightarrow
\text{local }U(3)\text{ covariance}
\Rightarrow
SU(3)\text{ as the non-Abelian part}.
\label{eq:threeToSU3}
\end{equation}
Thus $N=3$ does not constitute a mathematical proof that every three-state
system must be gauged by $SU(3)$.  It makes $SU(3)$ the minimal natural
non-Abelian gauge completion whenever local unitary basis redundancy is a
physical requirement.

\subsection{Color, left, and right carriers: a trinification-like target}
\label{subsec:333target}

Applying the same three-state gauge-completion principle to physically
distinct local carriers suggests
\begin{equation}
G_{333}=SU(3)_C\times SU(3)_L\times SU(3)_R.
\label{eq:G333new}
\end{equation}
The important logical order is that Eq.~\eqref{eq:G333new} is not imposed as
the primitive explanation of the number three.  Rather, the repeated
$SU(3)$ factors are candidate gauge realizations of repeated three-state
carriers inherited from the more primitive $N=3$ principle.  The present
framework should therefore be regarded as pointing toward a
trinification-like gauge completion, not as having derived a complete
ultraviolet trinification theory.

A useful schematic factorization of the internal space is
\begin{equation}
\mathcal H_{\rm total}
\simeq
\mathbb C_C^3
\otimes
(\mathbb C_{V,L}^3\oplus\mathbb C_{V,R}^3)
\otimes
\mathbb C_H^3.
\label{eq:totalfactornew}
\end{equation}
The first three-state factor carries color, the parity-paired vertical
factors carry the left/right high-scale gauge structure, and the final
factor carries the horizontal family spectrum.  The last factor is treated
differently, as explained below.

\subsection{Why color $SU(3)_C$ remains unbroken}
\label{subsec:colorunbrokennew}

The existence of several three-state carriers does not require them to
choose the same holonomy vacuum.  In the color sector we take the vacuum
holonomy to be completely degenerate,
\begin{equation}
\langle W_C\rangle=e^{i\theta_C}I_3,
\label{eq:colortrivialholonomy}
\end{equation}
so every color generator commutes with the vacuum,
\begin{equation}
[T_C^a,\langle W_C\rangle]=0,
\qquad a=1,\ldots,8.
\end{equation}
Before specifying its vacuum orientation, we clarify the status of the
vertical order parameter.  The field $\Sigma$ is not identified with the
Standard-Model Higgs and is not introduced as an additional elementary
Higgs multiplet.  It denotes an effective composite matrix order parameter
of the high-scale strong/internal dynamics.  Schematically it may arise from
fermion bilinears or other composite operators in a microscopic completion.
The present effective theory does not require a unique microscopic
realization.  The role of the strong dynamics is to generate a nonzero
condensate magnitude, while the role of the holonomy is to select its
orientation in the three-state carrier.

Accordingly, the vertical order parameter is taken to be a color singlet,
\begin{equation}
\langle\Sigma\rangle
=
I_C\otimes V_8T_8^{(V)}\otimes I_H,
\end{equation}
which implies
\begin{equation}
[T_C^a,\langle\Sigma\rangle]=0.
\end{equation}
Hence the vertical holonomy vacuum does not select a direction in color
space and
\begin{equation}
SU(3)_C\quad\text{remains unbroken}.
\label{eq:colorunbrokenresult}
\end{equation}
The difference between color and the vertical $L/R$ sectors is therefore
not the carrier dimension---all are three-dimensional---but the degeneracy
pattern of their vacuum holonomies.

\subsection{Why the horizontal three-state direction is not gauged}
\label{subsec:horizontalnotgauged}

The horizontal carrier also has dimension three, but its physical role is
different.  It is defined by a physical cyclic operator $C_H$ whose
eigenstates label the three fermion families.  We do not postulate an
arbitrary space-time dependent redundancy
\begin{equation}
\Psi_H(x)\longrightarrow G_H(x)\Psi_H(x)
\end{equation}
that would make all local horizontal bases physically equivalent.  Instead,
$C_H$ itself selects a preferred spectral basis.  Consequently,
\begin{equation}
\mathbb C_H^3
+\text{fixed cyclic holonomy}
\Rightarrow
\text{family spectral space, not a local }SU(3)_H\text{ gauge factor}.
\label{eq:Hnotgaugedboxed}
\end{equation}
This distinction prevents the mere occurrence of a three-state carrier from
being confused with automatic gauging.

Gauging the horizontal direction would introduce additional family gauge
bosons with interactions schematically of the form
\begin{equation}
g_H\,\overline\Psi_i\gamma^\mu(T_H^a)_{ij}\Psi_jA^a_{H\mu},
\end{equation}
which generically produce family-changing currents and would require an
additional breaking sector, anomaly analysis, and strong flavor
constraints.  None of these ingredients is required to realize the present
minimal family spectrum.  We therefore adopt the minimal distinction
\begin{equation}
\begin{array}{ccl}
\mathbb C_C^3 &:& \text{locally gauged }SU(3)_C,\\[1mm]
\mathbb C_{V,L}^3\oplus\mathbb C_{V,R}^3
&:& \text{locally gauged }SU(3)_L\times SU(3)_R,\\[1mm]
\mathbb C_H^3 &:& \text{non-gauged cyclic family holonomy}.
\end{array}
\label{eq:gaugedVsHorizontal}
\end{equation}
Thus the primitive number three is common, while gauging is determined by
whether local basis covariance is required in the corresponding carrier.

\section{$T_8$ splitting on both branches}
\label{sec:T8both}

Having identified the vertical three-state carriers with the local
$SU(3)_L\times SU(3)_R$ gauge factors of Sec.~\ref{sec:gaugeconnection},
the first high-scale splitting is taken to act similarly on both
three-state internal carriers.  Introduce
\begin{equation}
X_L=\Delta_L T_{8L},
\qquad
X_R=\Delta_R T_{8R},
\end{equation}
with
\begin{equation}
T_8=\frac{1}{2\sqrt3}\diag(1,1,-2).
\end{equation}

The $T_8$ direction can be related directly
to the primitive holonomy mass splitting
derived in Eq.~\eqref{eq:primitiveMassSplitting}.
At $\theta_3=0$, the three eigenvalues are
\begin{equation}
M_0+2\Delta,\qquad
M_0-\Delta,\qquad
M_0-\Delta.
\end{equation}
After ordering the degenerate pair as the
first two states, the corresponding diagonal
mass operator becomes
\begin{equation}
{\cal M}_{2+1}
=
M_0 I+
\Delta\diag(-1,-1,2).
\end{equation}
Using the standard normalization of $T_8$,
this can be written as
\begin{equation}
{\cal M}_{2+1}
=
M_0 I-2\sqrt3\,\Delta T_8.
\label{eq:holonomyT8Splitting}
\end{equation}

The $T_8$-type $2+1$ structure is therefore
already present in the eigenvalue spectrum
of the primitive three-state holonomy.
The twofold degeneracy identifies a
distinguished two-dimensional subspace,
while the remaining state is separated.

This spectral correspondence does not by
itself establish spontaneous gauge-symmetry
breaking. In the vertical sector, the
corresponding $T_8$ direction must be
realized by an appropriate dynamical order
parameter or symmetry-breaking vacuum.
The subsequent vacuum-selection mechanism
determines how this spectral structure
is inherited by the left- and right-handed
gauge sectors.
Equivalently, a $2+1$ mass pattern
\begin{equation}
M_{1}=M_{2}\neq M_3
\end{equation}
has a traceless component proportional to $T_8$.  Therefore
\begin{equation}
SU(3)_L\to SU(2)_L\times U(1)_L,
\qquad
SU(3)_R\to SU(2)_R\times U(1)_R.
\label{eq:T8breakingboth}
\end{equation}
At this stage the left and right branches are still structurally
symmetric.  In particular, a pure $T_8$ holonomy cannot by itself explain
why $SU(2)_L$ survives while $SU(2)_R$ disappears at a higher scale.
The distinction must arise from a second vacuum-selection mechanism.

Writing the holonomies in the Cartan subalgebra,
\begin{equation}
\langle W_L\rangle
=
\exp\left[i(\theta_{8L}T_{8L}+\phi_LT_{3L})\right],
\end{equation}
\begin{equation}
\langle W_R\rangle
=
\exp\left[i(\theta_{8R}T_{8R}+\phi_RT_{3R})\right],
\end{equation}
we see that
\begin{equation}
\phi_{L,R}=0
\end{equation}
leaves the first two eigenphases degenerate and therefore preserves the
corresponding $SU(2)$.  A nonzero $\phi_R$ lifts this degeneracy and
selects a direction inside $SU(2)_R$.

\section{Vertical and horizontal organization of the $3\times3$ fermion space}
\label{sec:VHorganization}

The preceding sections identified two parity-related internal branches,
$\mathbb C_L^3\oplus\mathbb C_R^3$, and showed how the vertical
three-state carrier can undergo a common $T_8$ splitting followed by a
right-handed vacuum selection.  We now make explicit how this chiral
structure is embedded in the $3\times3$ organization used for quarks and
leptons.

The two occurrences of ``three'' in this $3\times3$ notation have
different physical roles.  For each chirality branch $\chi=L,R$, the
vertical carrier is the three-state internal space on which the
$SU(3)_\chi$ connection and the high-scale holonomy act, while the
horizontal carrier labels the three-family cyclic structure.  Schematically,
\begin{equation}
\mathcal H_{\rm fermion}^{(\chi)}
\simeq
\mathbb C_{V,\chi}^{3}\otimes\mathbb C_H^{3},
\qquad \chi=L,R.
\label{eq:VHtensor}
\end{equation}
The vertical and horizontal holonomies therefore act on different tensor
factors and need not have the same angle or effective coefficients.
They are nevertheless assumed to inherit the same primitive three-body
cyclic principle rather than representing unrelated insertions of the
number three.

Suppressing the chirality label when no confusion arises, the quark and
lepton arrays may be written as
\begin{equation}
  \Psi_Q  =
  \tikz[baseline=(Q.base)]{
    \node[inner sep=0pt] (Q) {$
    \begin{pmatrix}
     u_1 & u_2 & u_3\\
     d_1 & d_2 & d_3\\
     D_1 & D_2 & D_3
    \end{pmatrix}$};

    \draw[<->]
      ([yshift=4mm]Q.north west)
      --
      node[midway,above]{\scriptsize Horizontal holonomy: family}
      ([yshift=4mm]Q.north east);

    \draw[<->]
      ([xshift=-4mm]Q.north west)
      --
      node[pos=-0.4,left=2mm,rotate=90]{\scriptsize Vertical holonomy}
      ([xshift=-4mm]Q.south west);
    },
  \qquad
  \Psi_\ell =
  \tikz[baseline=(L.base)]{
    \node[inner sep=0pt] (L) {$
    \begin{pmatrix}
     \nu_1 & \nu_2 & \nu_3\\
     e_1 & e_2 & e_3\\
     N_1 & N_2 & N_3
    \end{pmatrix}$};

    \draw[<->]
      ([yshift=4mm]L.north west)
      --
      node[midway,above]{\scriptsize Horizontal holonomy: family}
      ([yshift=4mm]L.north east);

    \draw[<->]
      ([xshift=-4mm]L.north west)
      --
      node[pos=-0.4,left=2mm,rotate=90]{\scriptsize Vertical holonomy}
      ([xshift=-4mm]L.south west);
  }.
\label{eq:3x3VH}
\end{equation}
For a fixed family $i$, the vertical triplets are therefore
\begin{equation}
\Psi_Q^{(i)}=
\begin{pmatrix}u_i\\ d_i\\ D_i\end{pmatrix},
\qquad
\Psi_\ell^{(i)}=
\begin{pmatrix}\nu_i\\ e_i\\ N_i\end{pmatrix}.
\label{eq:verticaltriplets}
\end{equation}
The first two vertical entries form the light pair, whereas the third
entry is the singlet-like heavy direction.  The horizontal index
$i=1,2,3$ labels the three-family structure.

\subsection{Vertical holonomy: the $2+1$ light--heavy decomposition}

The first vertical selection is governed by the $T_8$ pattern derived in
Sec.~\ref{sec:T8both}.  On either chirality branch,
\begin{equation}
W_{V,\chi}^{(8)}
=
\exp(i\theta_{8\chi}T_{8\chi}),
\qquad
X_{V,\chi}=\Delta_\chi T_{8\chi}.
\end{equation}
Since
\begin{equation}
T_8=\frac{1}{2\sqrt3}\diag(1,1,-2),
\end{equation}
the vertical carrier decomposes as
\begin{equation}
\mathbf 3_V\longrightarrow \mathbf 2_V\oplus\mathbf 1_V,
\label{eq:vertical21}
\end{equation}
so that
\begin{equation}
\begin{pmatrix}u_i\\d_i\\D_i\end{pmatrix}
\longrightarrow
\begin{pmatrix}u_i\\d_i\end{pmatrix}\oplus D_i,
\qquad
\begin{pmatrix}\nu_i\\e_i\\N_i\end{pmatrix}
\longrightarrow
\begin{pmatrix}\nu_i\\e_i\end{pmatrix}\oplus N_i.
\label{eq:verticaldecompfields}
\end{equation}
The essential point is that the $T_8$ holonomy does not distinguish
$u$ from $d$, nor $\nu$ from $e$.  It separates the third vertical
component from a degenerate pair.  Thus the high-scale vertical holonomy
selects the light--heavy organization before electroweak symmetry breaking.

We now make the two ingredients of this breaking explicit.  For each
chirality branch, $\Sigma_\chi$ is the corresponding effective composite
matrix order parameter.  Its nonzero magnitude is generated by the
underlying strong dynamics (for example through dimensional transmutation
and condensation in a microscopic completion), whereas the holonomy-induced
splitting $X_{V,\chi}=\Delta_\chi T_{8\chi}$ aligns the condensate along
the $T_{8\chi}$ direction.  Thus we write
\begin{equation}
\langle\Sigma_\chi\rangle=V_{8\chi}T_{8\chi},
\label{eq:SigmaV8definition}
\end{equation}
where $V_{8\chi}$ is the magnitude of the composite condensate and should
not be interpreted as a fundamental Higgs vacuum expectation value.  In
particular, the holonomy selects the direction of the condensate but does
not, by itself, determine its absolute magnitude.  With this distinction,
the singlet-like fermions acquire masses of order
\begin{equation}
M_{D_i}\sim y_{D_i}V_8,
\qquad
M_{N_i}\sim y_{N_i}V_8,
\label{eq:heavyDN}
\end{equation}
while broken vertical gauge directions have masses parametrically
$M_X\sim g_VV_8$.  The holonomy splitting $\Delta_\chi$, the order
parameter $V_{8\chi}$, and the physical heavy masses are therefore
conceptually distinct quantities.

The second, chiral selection discussed in Sec.~\ref{sec:heavyvac} acts
inside the right-handed doublet left over after Eq.~\eqref{eq:vertical21}.
A nonzero $\phi_R^{\rm vac}$ introduces a $T_{3R}$ component and lifts the
remaining right-handed degeneracy, whereas $\phi_L^{\rm vac}=0$ leaves
$SU(2)_L$ intact.  Thus the vertical sequence is
\begin{equation}
\mathbf 3_{V,L/R}
\xrightarrow{\;T_8\;}
\mathbf 2_{L/R}\oplus\mathbf 1_{L/R},
\qquad
\mathbf 2_R\xrightarrow{\;\phi_R^{\rm vac}\;}1_R\oplus1_R,
\qquad
\mathbf 2_L\ \text{preserved}.
\label{eq:verticalfullsequence}
\end{equation}
This makes clear that ``vertical'' refers to the high-scale internal
light--heavy and chiral organization, not to the family hierarchy.

\subsection{Separation from color}

The vertical and color carriers are both three-dimensional but belong to
different tensor factors.  As shown in
Sec.~\ref{subsec:colorunbrokennew}, the vertical holonomy and order
parameter are color singlets, while the color holonomy is fully degenerate,
$\langle W_C\rangle\propto I_3$.  Hence the $T_8$ direction selected in the
vertical sector does not break QCD color:
\begin{equation}
SU(3)_C\quad\text{remains unbroken}.
\end{equation}
This is an explicit example of the general principle that the same natural
three-state carrier dimension may support different low-energy symmetry
patterns depending on the vacuum holonomy eigenvalue degeneracy.

\section{Dynamical chiral holonomy vacuum from the existing vertical heavy sector}
\label{sec:heavyvac}

\subsection{Minimality principle: no additional spectator fermions}

The vertical $2+1$ decomposition already contains the heavy states needed
for the vacuum-selection dynamics.  We therefore adopt the minimality
principle
\begin{equation}
\text{no additional fermion is introduced solely to select the holonomy vacuum.}
\label{eq:minimalityprinciple}
\end{equation}
For each horizontal family $i$, the right-oriented vertical carriers are
\begin{equation}
\Psi_{Q,R}^{(i)}=
\begin{pmatrix}u_{R i}\\ d_{R i}\\ D_{R i}\end{pmatrix},
\qquad
\Psi_{\ell,R}^{(i)}=
\begin{pmatrix}\nu_{R i}\\ e_{R i}\\ N_{R i}\end{pmatrix},
\label{eq:Rverticaltriplets}
\end{equation}
and similarly on the left-oriented branch.  After the $T_8$ splitting,
the third components become the heavy singlet-like states $D_i$ and
$N_i$.  They are not new spectator fields: they are the third vertical
components already present in Eq.~\eqref{eq:3x3VH}.

The holonomy potential must nevertheless be generated by the \emph{full
vertical triplet}, not by an isolated $SU(2)_R$ singlet $D_i$ or $N_i$.
The reason is simple: after the $T_8$ splitting the third component is a
singlet under the residual $SU(2)_R$ and by itself cannot select a
$T_{3R}$ direction.  The relevant closed process therefore belongs to the
pre-hypercharge vertical carrier, in which all three components still
participate in the inherited $SU(3)_R$ internal dynamics.

\subsection{Closed vertical propagation and the one-winding term}

For a fixed family, consider the most general Hermitian three-link
vertical mass/transition matrix after field rephasings have concentrated
the gauge-invariant loop phase on one link,
\begin{equation}
\mathcal M_{V,R}^{(i)}(\Phi_R)=
\begin{pmatrix}
M_{1i} & t_{12,i} & t_{31,i}e^{-i\Phi_R}\\
t_{12,i} & M_{2i} & t_{23,i}\\
t_{31,i}e^{i\Phi_R} & t_{23,i} & M_{3i}
\end{pmatrix}.
\label{eq:verticalMRgeneral}
\end{equation}
For the quark carrier $M_{3i}$ evolves into the heavy $D_i$ scale, while
for the lepton carrier it evolves into the heavy $N_i$ scale.  Defining
\begin{equation}
A=m-M_{1i},\qquad B=m-M_{2i},\qquad C=m-M_{3i},
\end{equation}
the characteristic polynomial is
\begin{equation}
\det(mI-\mathcal M_{V,R}^{(i)})
=ABC-A t_{23,i}^{2}-B t_{31,i}^{2}-C t_{12,i}^{2}
-2t_{12,i}t_{23,i}t_{31,i}\cos\Phi_R.
\label{eq:verticalcharpoly}
\end{equation}
Thus the phase first appears only after completing the three-link closed
path
\begin{equation}
1\longrightarrow2\longrightarrow3\longrightarrow1,
\end{equation}
and its first contribution is
\begin{equation}
\Delta_{\rm wind}^{(1)}
\propto
t_{12,i}t_{23,i}t_{31,i}\cos\Phi_R.
\label{eq:onewindgeneral}
\end{equation}
In the symmetric limit
$M_{1i}=M_{2i}=M_{3i}=M$ and
$t_{12,i}=t_{23,i}=t_{31,i}=t$, Eq.~\eqref{eq:verticalcharpoly}
reduces to
\begin{equation}
(m-M)^3-3t^2(m-M)-2t^3\cos\Phi_R=0,
\label{eq:symmetricringlimit}
\end{equation}
recovering the characteristic $t^3\cos\Phi_R$ one-winding structure.
Two windings first generate a second harmonic of order
$t^6\cos2\Phi_R$ in this symmetric limit.  More generally,
\begin{equation}
V_{\rm hol}(\Phi_R)
=V_0+\sum_{n\ge1}c_n\,\mathcal T^{\,n}\cos(n\Phi_R),
\qquad
\mathcal T\sim
\frac{t_{12}t_{23}t_{31}}{M_{\rm heavy}^{3}},
\label{eq:windingexpDN}
\end{equation}
where the precise coefficients depend on the nondegenerate vertical
spectrum.

\subsection{The $D/N$ spectrum as the source of the effective potential}

The quark and lepton vertical spectra contribute to the same holonomy
potential.  Schematically,
\begin{align}
V_R(\phi_R)
={}&V_{Q,R}(\phi_R)+V_{\ell,R}(\phi_R)+\cdots,
\label{eq:VRDNsum}\\
V_{Q,R}
={}&-\frac{N_c}{16\pi^2}
\sum_{i=1}^{3}\sum_{a=1}^{3}
\bigl[m^{Q}_{ia}(\phi_R)\bigr]^4
\left[
\ln\frac{[m^{Q}_{ia}(\phi_R)]^2}{Q^2}-\frac32
\right],
\label{eq:VQDN}\\
V_{\ell,R}
={}&-\frac{1}{16\pi^2}
\sum_{i=1}^{3}\sum_{a=1}^{3}
\bigl[m^{\ell}_{ia}(\phi_R)\bigr]^4
\left[
\ln\frac{[m^{\ell}_{ia}(\phi_R)]^2}{Q^2}-\frac32
\right].
\label{eq:VlDN}
\end{align}
Here $m^{Q}_{ia}$ and $m^{\ell}_{ia}$ are the three eigenvalues of the
vertical quark and lepton carriers for horizontal family $i$.  The heavy
third components $D_i$ and $N_i$ give the dominant threshold sensitivity
when their $T_8$-generated masses are large.  The essential point is that
these are the same heavy states that later appear in the low-energy
light--heavy mixing matrices, There is no separate high-scale fermion
sector.

At low harmonic order one may write
\begin{equation}
V_R(\phi_R)
=a_R\cos\phi_R+b_R\cos2\phi_R+\cdots,
\label{eq:VRDNharmonics}
\end{equation}
with coefficients calculable, in a specified ultraviolet realization,
from the $D/N$-containing vertical spectra and transition amplitudes.

\subsection{Spontaneous left--right vacuum selection}

No explicit difference between the left and right particle content is
needed.  Before vacuum selection, parity requires a joint potential
symmetric under
\begin{equation}
\phi_L\longleftrightarrow\phi_R.
\end{equation}
A minimal parity-symmetric form illustrating spontaneous chiral selection
is
\begin{equation}
V_{LR}
=a(\cos\phi_L+\cos\phi_R)
+b(\cos2\phi_L+\cos2\phi_R)
+\kappa\cos\phi_L\cos\phi_R+\cdots.
\label{eq:VLRjoint}
\end{equation}
The mixed term is allowed because the constituent-mass structure already
connects the two parity-related branches.  It does not require an
additional field.

A stationary configuration with
\begin{equation}
\phi_L=0,
\qquad
\cos\phi_R^\star=-\frac{a+\kappa}{4b}
\label{eq:asymstationary}
\end{equation}
Here the superscript $\star$ denotes the stationary vacuum value of the
right-handed holonomy angle.  Thus $\phi_R^\star$ is a vacuum phase, not a
new scalar field and not the complex conjugate of a field $\Phi_R$.
It parametrizes the second, $T_{3R}$-directed vacuum selection inside the
$SU(2)_R$ doublet that remains after the $T_8$ splitting.

It exists when
\begin{equation}
|a+\kappa|<4b,
\qquad b>0.
\end{equation}
Its curvature in the nontrivial $R$ direction is
\begin{equation}
\left.
\frac{\partial^2V_{LR}}{\partial\phi_R^2}
\right|_{(0,\phi_R^\star)}
=4b\sin^2\phi_R^\star>0,
\end{equation}
while stability of the $L$ direction requires
\begin{equation}
a+4b+\kappa\cos\phi_R^\star<0.
\label{eq:Lstabilitycond}
\end{equation}
Whenever these conditions are satisfied, the parity-related pair
\begin{equation}
(\phi_L,\phi_R)=(0,\phi_\star),
\qquad
(\phi_\star,0)
\label{eq:paritypairvacua}
\end{equation}
forms a pair of asymmetric vacua.  Choosing the first vacuum gives
\begin{equation}
\phi_L^{\rm vac}=0,
\qquad
\phi_R^{\rm vac}=\phi_\star\neq0,
\label{eq:chiralvacuumDN}
\end{equation}
without introducing a new spectator fermion sector and without imposing
an explicit left--right asymmetry in the field content.
The holonomy angles $\theta_{8R}$ and $\phi_R$ are dimensionless variables 
specifying the orientation of the vacuum in the internal space, 
whereas the physical symmetry-breaking scale is set by the dimensionful 
condensate $V_{8R}$.

The vertical symmetry sequence is therefore
\begin{equation}
SU(3)_L\times SU(3)_R
\xrightarrow{\;T_8\;}
SU(2)_L\times SU(2)_R\times U(1)_L\times U(1)_R
\xrightarrow{\;\phi_R^\star\;}
SU(2)_L\times U(1)\times U(1).
\label{eq:breakingchainDN}
\end{equation}
Only after this high-scale internal selection do we identify the surviving
Abelian combination with hypercharge.  In the normalization used here,
\begin{equation}
Y=T_{3R}+\frac{T_{8L}+T_{8R}}{\sqrt3}.
\label{eq:YDN}
\end{equation}
This ordering is important: the closed vertical transitions in
Eq.~\eqref{eq:verticalMRgeneral} belong to the pre-hypercharge internal
$SU(3)_R$ dynamics and should not be interpreted as low-energy mass
mixing between fields of already fixed electromagnetic charge.
More precisely, $V_{8R}$ provides the dimensionful radial scale 
of the right-handed composite order parameter, whereas $\phi_R$ 
specifies its orientation within the residual $SU(2)_R$ sector. 
The nonzero stationary value $\phi_R^\star$ therefore does not introduce 
a second independent mass scale.
Rather, it converts the already generated condensate scale into masses 
for the gauge bosons associated with the broken right-handed directions, 
schematically $M_{W_R}\sim g_R V_{8R}F(\phi_R^\star)$, with $F(0)=0$.

\section{Universal horizontal holonomy and the three-family backbone}
\label{sec:horizontal}

The horizontal direction acts on the column index of
Eq.~\eqref{eq:3x3VH}.  Unlike the vertical holonomy, it does not separate
a light pair from a heavy singlet.  Its role is to provide a common
threefold spectral backbone for the three observed families.

Although this carrier is also $\mathbb C^3$, it is not gauged in the
minimal construction.  The physical cyclic operator $C_H$ selects a family
eigenbasis, so horizontal transformations are not treated as arbitrary
space-time dependent redundancies.  This is the distinction established in
Sec.~\ref{subsec:horizontalnotgauged}: three states make an $SU(3)$ gauge
completion natural only when local unitary basis covariance is required.
Keeping the horizontal sector spectral rather than gauged also avoids
introducing unnecessary family gauge bosons and their associated
flavor-changing currents.

The minimal horizontal cyclic operator is
\begin{equation}
C_H(\theta_H)=
\begin{pmatrix}
0&1&0\\
0&0&1\\
e^{i\theta_H}&0&0
\end{pmatrix},
\qquad
C_H^3=e^{i\theta_H}I.
\label{eq:CHmatrix}
\end{equation}
Its three eigenvalues are
\begin{equation}
\lambda_k=e^{i\varphi_k},
\qquad
\varphi_k=\frac{\theta_H+2\pi k}{3},
\qquad k=0,1,2,
\label{eq:Hphases}
\end{equation}
with Fourier-type eigenvectors
\begin{equation}
|k\rangle_H
=\frac{1}{\sqrt3}
\begin{pmatrix}
1\\ e^{i\varphi_k}\\ e^{2i\varphi_k}
\end{pmatrix}.
\label{eq:Heigenvectors}
\end{equation}
A Hermitian family operator built from the first two nontrivial cyclic
harmonics may be written as
\begin{equation}
H_H(\theta_H;r,s)
=
I+r(C_H+C_H^\dagger)
+s(C_H^2+C_H^{\dagger2}).
\label{eq:HH}
\end{equation}
Because $H_H$ is a polynomial in $C_H$ and $C_H^\dagger$, it has the same
eigenvectors, and its eigenvalues are
\begin{equation}
h_k
=
1+2r\cos\varphi_k+2s\cos2\varphi_k,
\qquad k=0,1,2.
\label{eq:familyeigs}
\end{equation}

The corresponding three-family mass
spectrum follows directly from these
holonomy eigenvalues. For the fermion
sector $f$, we obtain
\begin{equation}
M_f^{[k]}
=
M_{0f}
+2\Delta_f\cos\varphi_k
+2\Delta_f^{(2)}\cos2\varphi_k,
\qquad k=0,1,2,
\label{eq:horizontalMassSplitting}
\end{equation}
where
\begin{equation}
M_{0f}=m_f^{(0)},\qquad
\Delta_f=m_f^{(0)}r_f,\qquad
\Delta_f^{(2)}=m_f^{(0)}s_f.
\end{equation}

The three-body holonomy
mass relation
\begin{equation}
M_f^{[k]}
=
M_{0f}
+2\Delta_f
\cos\left(
\frac{\theta_H+2\pi k}{3}
\right)++2\Delta_f^{(2)}
\cos\left(
\frac{2\theta_H+4\pi k}{3}
\right).
\label{eq:horizontalLeadingMass}
\end{equation}

The universal horizontal holonomy
therefore provides a common threefold
spectral structure for quarks and leptons,
while the sector-dependent parameters
$M_{0f}$, $\Delta_f$, and
$\Delta_f^{(2)}$ determine their
mass scales and splittings.

At $\theta_H=0$, the leading-harmonic
spectrum exhibits a twofold degeneracy.
A nonzero holonomy phase, higher cyclic
harmonics, and heavy-fermion mixing
can lift this degeneracy and contribute
to the observed mass hierarchies.

The quantities $M_f^{[k]}$ are eigenvalues
of the unperturbed Hermitian texture.
Physical fermion masses are obtained
from the singular values of the full
effective mass matrices, including
the heavy-sector corrections discussed
in the following section.
The nominal third harmonic carries no new family dependence because
\begin{equation}
\cos3\varphi_k
=
\cos(\theta_H+2\pi k)
=
\cos\theta_H.
\end{equation}
Hence the first two harmonics form the minimal nontrivial completion of
the leading single-cosine spectrum.

The phrase \emph{universal horizontal holonomy} means that the same
cyclic geometry and, in the minimal limit, the same phase $\theta_H$ are
inherited by the quark and lepton sectors.  It does not require identical
coefficients in every sector.  We therefore write
\begin{equation}
M_f^{(0)}
=
m_f^{(0)}
H_H(\theta_H;r_f,s_f),
\qquad
f=u,d,e,\nu,
\label{eq:universalHsector}
\end{equation}
where $r_f$ and $s_f$ encode sector-dependent couplings to the common
horizontal geometry.  To make the starting point of the numerical fit
explicit, Eq.~\eqref{eq:universalHsector} can be written in matrix form as
\begin{equation}
M_f^{(0)}=m_f^{(0)}
\begin{pmatrix}
1 & r_f+s_f e^{-i\theta_H} & r_f e^{-i\theta_H}+s_f\\
r_f+s_f e^{i\theta_H} & 1 & r_f+s_f e^{-i\theta_H}\\
r_f e^{i\theta_H}+s_f & r_f+s_f e^{i\theta_H} & 1
\end{pmatrix},
\label{eq:M0matrixExplicit}
\end{equation}
for the convention of Eq.~\eqref{eq:CHmatrix}.  This display is not an
additional ansatz, It is simply the polynomial $H_H$ written out explicitly.
The independent holonomy-backbone parameters entering a sector are therefore
$m_f^{(0)}$, $r_f$, $s_f$, and the common phase $\theta_H$.

In the absence of vertical deformation, matrices
that are only functions of the same $C_H$ commute and are simultaneously
diagonalizable.  Nontrivial CKM and PMNS mixing therefore requires a
misalignment or deformation supplied by the vertical/heavy sector.

This distinction is central:
\begin{equation}
\begin{aligned}
\text{vertical holonomy}&:\quad
3_V\to2_V+1_V,
\ \text{heavy states and chiral selection},\\
\text{horizontal holonomy}&:\quad
k=0,1,2,
\ \text{family replication and hierarchy}.
\end{aligned}
\label{eq:VHroles}
\end{equation}
The two directions meet only when the physical fermion mass matrices are
formed.

\section{Physical fermion masses from the vertical--horizontal interplay}
\label{sec:physicalmasses}

The low-energy fermion spectrum is generated only after the vertical and
horizontal structures are combined with electroweak symmetry breaking.
The horizontal holonomy supplies a common family basis and a bare
threefold spectrum. The vertical sector supplies heavy partners and a
source of misalignment, and the electroweak Higgs couples the surviving
left- and right-handed light states.

\subsection{Electroweak breaking as the left--right mass bridge}

After the high-scale holonomy breakings, the surviving electroweak group
contains $SU(2)_L\times U(1)_Y$.  With
\begin{equation}
\langle H_{\rm EW}\rangle
=\frac{1}{\sqrt2}
\begin{pmatrix}0\\v\end{pmatrix},
\end{equation}
the ordinary Dirac terms take the form
\begin{equation}
\mathcal L_m
\supset
-\overline\Psi_L M_f^{(0)}\Psi_R+\text{h.c.}
\end{equation}
This is conceptually distinct from the high-scale vertical breaking:
\begin{equation}
\begin{aligned}
\text{vertical holonomy}&:\quad 3\to2+1\ \text{and chiral selection},\\
\text{EW breaking}&:\quad L\leftrightarrow R\ \text{mass generation}.
\end{aligned}
\end{equation}
In the present framework, the electroweak Higgs field $H$ is kept as the ordinary 
$SU(2)_L$ doublet of the Standard Model. 
It should be distinguished from the composite order parameter $\Sigma$, 
which is associated with the higher-scale $SU(3)\to SU(2)\times U(1)$ breaking. 
Thus the present construction does not assume 
that the Standard-Model Higgs is itself a fundamental $SU(3)$ triplet. 
A possible embedding of $H$ into a parent $SU(3)_L$ multiplet, for example 
through the decomposition $\mathbf 3\to\mathbf 2\oplus\mathbf 1$, 
is an interesting extension 
but is not required for the minimal framework considered here.
The origin of the electroweak Higgs doublet is not addressed 
in the present effective construction.

\subsection{Up-type quarks as a horizontal reference sector}

A minimal reference realization is to let the up-type quarks follow the
horizontal spectrum directly,
\begin{equation}
M_u=m_u^{(0)}H_H(\theta_H;r_u,s_u).
\label{eq:MuH}
\end{equation}
Its eigenvalues are therefore proportional to
$h_k(\theta_H;r_u,s_u)$.  The up sector then defines a convenient
horizontal basis against which vertical deformations in the down and
neutral-lepton sectors can be measured.

As a numerical illustration, we restrict Eq.~\eqref{eq:familyeigs}
and denote
$M\equiv m_u^{(0)}$, $\Delta\equiv m_u^{(0)}r_u$, $\Delta^{(2)}\equiv m_u^{(0)}s_u$ and
$S\equiv\theta_H$. The three signed eigenvalues in this reference
limit can be written, with a convenient ordering, as
\begin{align}
\lambda_1 &= M+2\Delta\cos(S/3)+2\Delta^{(2)}\cos(2S/3),\nonumber\\
\lambda_2 &= M-\Delta\cos(S/3)-\sqrt{3}\Delta\sin(S/3)
              -\Delta^{(2)}\cos(2S/3)+\sqrt{3}\Delta^{(2)}\sin(2S/3),\nonumber\\
\lambda_3 &= M-\Delta\cos(S/3)+\sqrt{3}\Delta\sin(S/3)
              -\Delta^{(2)}\cos(2S/3)-\sqrt{3}\Delta^{(2)}\sin(2S/3).
\label{eq:upillustrativeeigenvalues}
\end{align}
For definiteness, take the illustrative positive eigenvalue assignment
$(\lambda_1,\lambda_2,\lambda_3)=(m_t,m_u,m_c)$, with masses in GeV,
\begin{equation}
(m_t,m_c,m_u)=(170,\ 0.630,\ 0.0012),
\label{eq:upillustrativeinput}
\end{equation}
approximating the $\overline{\mathrm{MS}}$ running masses at the common
scale $\mu=M_Z$ inferred from the Standard Model Yukawa couplings and
running vacuum expectation value reported in Ref.~\cite{AntuschHinzeSaad2026}.
These rounded inputs are intended for illustration, not for a precision
fit.  The corresponding scheme-consistent relation is
$m_q(M_Z)=y_q(M_Z)v(M_Z)/\sqrt{2}$.

The sum and difference of Eq.~\eqref{eq:upillustrativeeigenvalues}
give particularly transparent constraints:
\begin{align}
3M &= m_t+m_c+m_u=170\ \mathrm{GeV},\nonumber\\
2\bigl[M-\Delta\cos(S/3)-\Delta^{(2)}\cos(2S/3)\bigr]&=m_c+m_u=0.631\ \mathrm{GeV},\nonumber\\
2\sqrt{3}\Delta\sin(S/3)-2\sqrt{3}\Delta^{(2)}\sin(2S/3)&=m_c-m_u=0.628\ \mathrm{GeV}.
\label{eq:upillustrativeconstraints}
\end{align}
Here we assume $\Delta =\Delta^{(2)}$, then 
\begin{align}
M&=56.9\ \mathrm{GeV},\nonumber\\
\Delta\cos(S/3)+\Delta\cos(2S/3)&=56.3\ \mathrm{GeV},\nonumber\\
\Delta\sin(S/3)-\Delta\sin(2S/3)&=0.18\ \mathrm{GeV},
\end{align}
which yields
\begin{equation}
\Delta\simeq28.3\ \mathrm{GeV},\qquad
S\simeq-0.0193.
\label{eq:upillustrativefit}
\end{equation}
Thus, a common mass parameter and a cyclic mixing parameter of
comparable magnitude produce one heavy eigenvalue and two much lighter
ones. For small $S$, the spectrum takes the simple form
\begin{equation}
\lambda_1\simeq M+4\Delta,\qquad
\lambda_{2,3}\simeq M-2\Delta\pm\frac{\Delta S}{\sqrt3},
\label{eq:upsmallphase}
\end{equation}
up to corrections of order $S^2$. The near-degeneracy $M\simeq2\Delta$
places the two light states near zero, while the cyclic phase splits
them. Both parameters may arise from electroweak-symmetry-breaking
Yukawa terms, so their comparable scales are plausible. Equality is
not, however, implied by a common Higgs vacuum expectation value.
Moreover, the very small up-quark mass requires a near cancellation
between $M-\Delta\cos(S/3)$ and
$\sqrt3\Delta\sin(S/3)$. Therefore this exact three-input,
three-parameter reproduction is an illustrative consistency check,
not a prediction or an explanation of flavor naturalness. Possible
vertical/right-handed mixing and higher cyclic harmonics can modify
this leading-limit fit, while the common phase $S=\theta_H$ remains
subject to constraints from the other fermion sectors.

\subsection{Down-type quarks and heavy $D$ mixing}

Before heavy mixing, the down-type states may inherit the same horizontal
geometry,
\begin{equation}
M_d^{(0)}=m_d^{(0)}H_H(\theta_H;r_d,s_d).
\end{equation}
The vertical breaking simultaneously generates heavy singlet-like
partners $D_i$ with masses $M_{D_i}\sim y_{D_i}V_8$.  The complete
light--heavy system is therefore
\begin{equation}
\mathcal M_{dD}
=
\begin{pmatrix}
M_d^{(0)}&X_D\\
Y_D&M_D
\end{pmatrix}.
\label{eq:MdDfullnew}
\end{equation}
For $M_D$ larger than the electroweak entries, integrating out the heavy
states gives
\begin{equation}
M_d^{\rm eff}
\simeq
M_d^{(0)}-X_DM_D^{-1}Y_D.
\label{eq:downseesaw}
\end{equation}
Thus the physical down spectrum is a horizontal spectrum deformed by the
vertical heavy sector.

If the common holonomy approximately aligns the relevant matrices, the
third-family $b$--$D$ block can be written schematically as
\begin{equation}
\mathcal M_{bD}
=
\begin{pmatrix}
m_b^{(0)}&\mu_L\\
\mu_R&M_D
\end{pmatrix}.
\end{equation}
A small physical singular value is naturally obtained near the
rank-one condition
\begin{equation}
\mu_L\mu_R\simeq m_b^{(0)}M_D.
\label{eq:rankonebD}
\end{equation}
In this interpretation the large top--bottom splitting need not arise
entirely from unrelated fundamental horizontal Yukawa parameters: the
up sector can remain close to the direct horizontal spectrum while the
down sector is substantially deformed by vertical mixing.  Equation
\eqref{eq:rankonebD} is a structural possibility rather than, at the
present stage, a derived numerical prediction.

\subsection{From fit parameters to the CKM matrix}
\label{subsec:CKMpropagation}

The fitting procedure is most transparent when the holonomy parameters and
the heavy-sector parameters are kept conceptually separate.  We write
\begin{equation}
M_u^{\rm eff}=M_u^{(0)}+\delta M_u,
\qquad
M_d^{\rm eff}=M_d^{(0)}+\delta M_d^{D},
\qquad
\delta M_d^{D}=-X_DM_D^{-1}Y_D.
\label{eq:MuMdFitDefinition}
\end{equation}
The zeroth-order matrices determine the principal mass hierarchy through
$\{m_{0q},r_q,s_q,\theta_H\}$, whereas the noncommuting parts of
$\delta M_u$ and $\delta M_d^D$ generate the weak--mass-basis
misalignment.

Define
\begin{equation}
{\cal H}_q=M_q^{\rm eff}M_q^{{\rm eff}\dagger}.
\end{equation}
The left diagonalization matrices satisfy
\begin{equation}
U_{qL}^\dagger{\cal H}_qU_{qL}
=\diag(m_{q1}^2,m_{q2}^2,m_{q3}^2),
\end{equation}
and the CKM matrix follows from
\begin{equation}
V_{\rm CKM}=U_{uL}^\dagger U_{dL}.
\end{equation}
Because $M_u^{(0)}$ and $M_d^{(0)}$ are functions of the same $C_H$, they
share a Fourier-type zeroth-order eigenbasis $F_H$.  Hence
\begin{equation}
V_{\rm CKM}^{(0)}=F_H^\dagger F_H=I.
\label{eq:CKMzero}
\end{equation}
The nontrivial CKM structure is therefore not put into the zeroth-order
holonomy matrix by hand. It is generated by the difference between the
up- and down-sector deformations.

For small deformations let
\begin{equation}
\widetilde{\delta{\cal H}}_q
=F_H^\dagger\delta{\cal H}_qF_H,
\qquad
\delta{\cal H}_q
=M_q^{(0)}\delta M_q^\dagger
+\delta M_qM_q^{(0)\dagger}
+\cdots.
\end{equation}
First-order perturbation theory gives
\begin{equation}
(\Omega_q)_{ij}
\simeq
\frac{(\widetilde{\delta{\cal H}}_q)_{ij}}
{m_{qj}^{(0)2}-m_{qi}^{(0)2}},
\qquad i\ne j,
\label{eq:OmegaFit}
\end{equation}
so that
\begin{equation}
V_{\rm CKM}\simeq I+\Omega_d-\Omega_u.
\label{eq:CKMfirstorder}
\end{equation}
Equation~\eqref{eq:CKMfirstorder} is the desired parameter-propagation map.
In a down-dominated deformation, the transformed entries
$(\widetilde{\delta{\cal H}}_d)_{12}$,
$(\widetilde{\delta{\cal H}}_d)_{23}$, and
$(\widetilde{\delta{\cal H}}_d)_{13}$ dominantly control
$V_{us}$, $V_{cb}$, and $V_{ub}$, respectively, although the Fourier
rotation means that the correspondence is not exactly one-to-one.

The physical CKM phase is generated only by a rephasing-invariant relative
complex structure.  If
\begin{equation}
(\delta M_q)_{ij}=|\epsilon_{ij}^{(q)}|e^{i\phi_{ij}^{(q)}},
\end{equation}
then phases that can be removed by quark-field redefinitions are
unobservable, while the remaining relative phases\cite{KobayashiMaskawa1973, Jarlskog1985} enter
\begin{equation}
J_{\rm CKM}
=\operatorname{Im}
\left(V_{us}V_{cb}V_{ub}^*V_{cs}^*\right).
\end{equation}
Thus the immediate origin of $\delta_{\rm CKM}$ is the complex
misalignment between the common holonomy backbone and the heavy-induced
flavor deformation.  In the most predictive version of the framework,
these deformation phases should be functions of $\theta_H$ or of the
underlying three-body holonomy rather than independent fit parameters.

For numerical work we therefore distinguish the parameter sets
\begin{equation}
\begin{aligned}
P_{\rm hol}&=\{m_{0u},r_u,s_u,m_{0d},r_d,s_d,\theta_H\},\\
P_{\rm heavy}&=\{M_D,(X_D)_{ij},(Y_D)_{ij},\ldots\},\\
O_{\rm fit}&=\{m_u,m_c,m_t,m_d,m_s,m_b,
|V_{us}|,|V_{cb}|,|V_{ub}|,\delta_{\rm CKM}\}.
\end{aligned}
\label{eq:FitParameterMap}
\end{equation}
A useful diagnostic of predictivity is the sensitivity matrix
\begin{equation}
S_{ia}=\frac{\partial O_i}{\partial p_a},
\qquad p_a\in P_{\rm hol}\cup P_{\rm heavy},
\label{eq:sensitivitymatrix}
\end{equation}
which makes explicit which fitted observables are controlled by each
underlying parameter.

The explicit matrix-level realization, including the heavy-state
resolvent and the non-unitary charged-current correction, is given in
Appendix~\ref{app:weak-mass-transmission}.  It uses the same
$C_H(\theta_H)$ as the horizontal backbone, rather than introducing a
second, unrelated cyclic phase.  Its restricted diagonal mixing blocks
are an illustrative weak-basis texture, not a consequence already
established by the ultraviolet theory.

\subsection{Charged leptons and the neutral vertical sector}

Without a charged heavy partner, the charged-lepton matrix can provide a
comparatively direct probe of the horizontal geometry,
\begin{equation}
M_e=m_e^{(0)}H_H(\theta_H;r_e,s_e).
\label{eq:MeH}
\end{equation}
The neutral sector is different because the vertical singlet-like states
$N_i$ are available.  A generic block form is
\begin{equation}
\mathcal M_{\nu N}
=
\begin{pmatrix}
M_\nu^{(0)}&X_N\\
Y_N&M_N
\end{pmatrix}.
\label{eq:MnuN}
\end{equation}
For a Majorana-type heavy sector this reduces schematically to
\begin{equation}
M_\nu^{\rm eff}
\simeq
-X_NM_N^{-1}X_N^T,
\label{eq:Mnuseesaw}
\end{equation}
whereas the precise formula differs for a purely Dirac realization.
Because $M_N$ belongs to the vertical sector, $M_\nu^{\rm eff}$ need not
be diagonalized by the same horizontal Fourier basis as $M_e$.  Hence
\begin{equation}
U_{\rm PMNS}=U_e^\dagger U_\nu
\end{equation}
can be nontrivial even when the charged-lepton backbone is controlled by
the same $C_H$.  The same logic applies to CKM mixing: if both up and
down matrices were functions only of $C_H$, they would commute. Vertical
heavy-state deformation supplies the required noncommuting structure.

\subsection{Leptonic CP violation from the common three-body holonomy}
\label{subsec:leptonCP}

The horizontal three-state carrier is common to the quark and lepton
sectors.  We therefore use the same holonomy angle $\theta_H$ rather
than introducing a separate leptonic phase $\phi$.  The charged-lepton
and neutral-lepton sectors nevertheless need not yield identical
observable CP phases. The latter is affected by the vertical heavy
neutral states and the resulting misalignment of mass eigenbases.
In this subsection we do not impose an additional $\mu$--$\tau$
reflection symmetry.

Let $C$ be the three-state cyclic shift, with $C^3=I_3$ and
$C^\dagger=C^2$, and define
\begin{align}
 W_H(\theta_H)&=e^{i\theta_H/3}C,\\
 H_1(\theta_H)&=W_H+W_H^\dagger,\\
 H_2(\theta_H)&=W_H^2+(W_H^\dagger)^2.
 \label{eq:lepton-horizontal-operators}
\end{align}
Since $W_H$ is unitary, $H_2=H_1^2-2I_3$.  This identity relates
the one- and two-winding operators, but it does not fix the ratio
of their coefficients. That ratio depends on the dynamics generating
the effective interactions.  In particular, the two coefficients below
are phenomenological real parameters, not independently established
ultraviolet predictions.

A charged-lepton horizontal reference texture can be parameterized as
\begin{equation}
 M_e=m_e^{(0)}J_3+t_{1e}H_1(\theta_H)+t_{2e}H_2(\theta_H),
 \qquad (J_3)_{ij}=1,
 \label{eq:lepton-charged-reference}
\end{equation}
where all coefficients are real.  This democratic-plus-cyclic
form is an illustrative effective texture, not a derived consequence
of the six-fermion operator.  It is diagonalized biunitarily,
$U_{eL}^\dagger M_e U_{eR}=\operatorname{diag}(m_e,m_\mu,m_\tau)$.
For the Hermitian form in Eq.~\eqref{eq:lepton-charged-reference},
left and right rotations may be chosen equal up to eigenvalue-sign
conventions.  Fixing the three charged-lepton masses alone does not
uniquely determine $\theta_H$ if $m_e^{(0)},t_{1e},t_{2e}$ are all free.

The vertical sector introduces heavy neutral states $N_i$.  To
illustrate how the same horizontal phase can reach light-neutrino
observables, consider a Majorana-type block mass matrix,
\begin{equation}
 \mathcal M_{\nu N}=\begin{pmatrix}M_L&X_N\\X_N^T&M_N\end{pmatrix},
 \qquad M_N=M_N^T,
 \label{eq:lepton-heavy-block}
\end{equation}
with the minimal Type-I limit $M_L=0$.  One possible restricted
weak-basis ansatz that introduces no independent complex coupling is
\begin{align}
 X_N&=D_X+t_{1N}H_1(\theta_H)+t_{2N}H_2(\theta_H),\\
 D_X&=\operatorname{diag}(x_1,x_2,x_3),\qquad
 M_N=\operatorname{diag}(M_1,M_2,M_3),
 \label{eq:lepton-heavy-texture}
\end{align}
where $x_i,t_{1N},t_{2N},M_i$ are real, and $M_i$ are nonzero.
The non-universal real diagonal entries summarize a possible vertical
splitting. Their detailed origin and correlations with the charged
sector require further dynamical matching.  The ansatz of
Eq.~\eqref{eq:lepton-heavy-texture} should not be mistaken for a
unique consequence of the primitive holonomy.

When the heavy eigenvalues are large compared with the mixing blocks,
integration of the heavy states gives
\begin{equation}
 M_\nu^{\mathrm{eff}}
 \simeq M_L-X_NM_N^{-1}X_N^T.
 \label{eq:lepton-seesaw-common-phase}
\end{equation}
Thus the orientation-dependent terms $e^{\pm i\theta_H/3}$ and
$e^{\pm2i\theta_H/3}$ in $X_N$ propagate into the effective
complex symmetric neutrino mass matrix.  The physical mismatch is
obtained only after diagonalizing \emph{both} lepton sectors:
\begin{align}
 U_\nu^T M_\nu^{\mathrm{eff}}U_\nu
 &=\operatorname{diag}(m_1,m_2,m_3),\\
 U_{\mathrm{PMNS}}&=U_{eL}^\dagger U_\nu.
 \label{eq:lepton-PMNS-common-phase}
\end{align}
The leptonic Dirac CP phase is encoded in the rephasing-invariant
Jarlskog quantity
\begin{equation}
 J_{\mathrm{CP}}=\operatorname{Im}\!\left[
 (U_{\mathrm{PMNS}})_{e1}(U_{\mathrm{PMNS}})_{\mu2}
 (U_{\mathrm{PMNS}})^*_{e2}(U_{\mathrm{PMNS}})^*_{\mu1}
 \right]
 =c_{12}s_{12}c_{23}s_{23}c_{13}^{2}s_{13}\sin\delta_{\mathrm{CP}}.
 \label{eq:lepton-JCP-common-phase}
\end{equation}
Consequently $\delta_{\mathrm{CP}}$ is not identical to $\theta_H$.
It depends on the real vertical and heavy-sector parameters as well
as their interplay with charged-lepton diagonalization.

This construction has a transparent CP-conserving limit.  With all
other parameters real, at $\theta_H=0$ both $H_1$ and $H_2$ are real,
and hence $M_e$, $X_N$, $M_N$, and $M_\nu^{\mathrm{eff}}$ can all be
chosen real.  Thus physical Dirac-type CP violation vanishes in this
limit.  More generally, complex conjugation under
$\theta_H\mapsto-\theta_H$ implies CP-odd invariants change sign
when all other real parameters and mass-state identification are held
fixed:
\begin{equation}
 J_{\mathrm{CP}}(-\theta_H)=-J_{\mathrm{CP}}(\theta_H).
 \label{eq:lepton-CP-oddness}
\end{equation}
At special degeneracies, mixing angles and the convention-dependent
$\delta_{\mathrm{CP}}$ itself need not be uniquely defined. The
CP-conserving statement concerns physical invariants.

For illustration, a phenomenological exploration may hold the
quark-sector value $\theta_H\simeq-0.0193$ fixed, use measured
charged-lepton masses to constrain Eq.~\eqref{eq:lepton-charged-reference},
and then fit the neutrino mass-squared splittings and mixing angles
through Eq.~\eqref{eq:lepton-seesaw-common-phase}.  Such a procedure
tests whether the shared phase admits a viable parameter region, but
it does not yet constitute a parameter-independent prediction for
$\delta_{\mathrm{CP}}$.  In particular, the absolute neutrino mass
scale, the full charged-lepton rotation, and the sensitivity to
cancellations between one- and two-winding terms must be checked
before quoting a numerical CP prediction.

An exact $\mu$--$\tau$ reflection ansatz would yield
$\theta_{23}=45^\circ$ and $\delta_{\mathrm{CP}}=\pm90^\circ$ under
standard nondegeneracy and nonzero-angle conditions, but these are
predictions of that additional symmetry, not generic
consequences of the three-body holonomy.  We therefore do not use
such a restriction in the present minimal construction.

\subsection{One heavy sector, two energy-regime roles}

No separate high-scale spectator fermions are introduced.  The same
vertical heavy states $D_i$ and $N_i$ appear in two different energy
regimes:
\begin{enumerate}
\item at the high scale, they are the heavy third components of the full
vertical triplets, and the closed three-link propagation of those
triplets contributes to the holonomy effective potential that selects
the chiral vacuum
\item after the high-scale vacuum has been selected and electroweak
symmetry is broken, the surviving heavy states $D_i$ and $N_i$ can mix
with the light fermions and deform the universal horizontal family
spectrum.
\end{enumerate}
Thus the distinction is one of \emph{role and scale}, not one of particle
content.  The model uses one vertical heavy sector to connect high-scale
vacuum selection with low-energy flavor phenomenology.

\section{Hierarchy of scales and sequential breaking}
\label{sec:scales}

The combined vertical--horizontal picture naturally separates several
physical scales,
\begin{align}
\Lambda_3 &: \quad
\text{primitive three-body holonomy and selection of }N=3,
\\
\Lambda_8 &: \quad
\text{vertical }T_8\text{ selection and }3_V\to2_V+1_V,
\\
\Lambda_R &: \quad
\text{$D/N$-assisted vertical holonomy vacuum selection inside }SU(2)_R,
\\
M_{D,N} &: \quad
\text{vertical heavy-partner masses of order }yV_8,
\\
v &: \quad
SU(2)_L\times U(1)_Y\to U(1)_{\rm em}.
\end{align}
We require parametrically
\begin{equation}
\Lambda_3\gtrsim\Lambda_8\gtrsim\Lambda_R\gg v,
\end{equation}
with $M_{D,N}$ tied to the vertical-breaking regime but not necessarily
identical to $\Lambda_8$ or $\Lambda_R$.

The complete sequence can then be summarized as
\begin{equation}
\begin{aligned}
&\text{primitive three-body holonomy}
\Rightarrow \mathbb C_{V,L}^3\oplus\mathbb C_{V,R}^3,
\\
&\text{vertical }T_8\text{ holonomy}
\Rightarrow 3_V\to2_V+1_V
\Rightarrow D_i,N_i\ \text{heavy},
\\
&\text{right-holonomy vacuum}
\Rightarrow SU(2)_R\to U(1)_R,
\qquad SU(2)_L\ \text{preserved},
\\
&\text{horizontal holonomy}
\Rightarrow k=0,1,2\ \text{family backbone},
\\
&\text{EW breaking + vertical heavy mixing}
\Rightarrow \text{physical quark and lepton masses and mixings}.
\end{aligned}
\label{eq:fullVHsequence}
\end{equation}
The horizontal direction is therefore not an additional stage of gauge
symmetry breaking.  It persists as the family backbone across the
vertical sequence and becomes observable through the low-energy mass
spectrum.

\section{Phenomenological consequences}

The low-energy theory should approach the Standard Model when all heavy
states and extra gauge bosons are sufficiently decoupled.  
Residual mixing generates characteristic deviations,
including violations of CKM unitarity and
flavor-changing neutral currents, as discussed
in the general framework of vector-like
singlet quarks~\cite{Branco2024}.
For heavy vector-like down states,
\begin{equation}
\Theta_D=X_DM_D^{-1},
\end{equation}
leading to
\begin{equation}
V_{\rm CKM}^\dagger V_{\rm CKM}
=I-\Theta_D\Theta_D^\dagger+\cdots,
\end{equation}
and to flavor-changing neutral-current effects controlled by
$\Theta_D\Theta_D^\dagger$.

The high-scale chiral vacuum implies additional heavy states associated
with the broken $SU(2)_R$ directions.  Their precise mass scale depends
on the ultraviolet realization, but parametrically
\begin{equation}
M_{W_R}\sim g_R\Lambda_R.
\end{equation}
The framework therefore predicts that the low-energy chiral structure is
accompanied by a high-scale right-handed sector, even if it is beyond
direct collider reach.

\section{What is derived, what is inherited, and what remains open}

For clarity, the status of the main ingredients is as follows.

\paragraph{Derived within the low-dimensional three-body framework.}
The operator $Q_3$, the two massless sectors $Q_3=\pm1$, their opposite
holonomy phases, parity exchange of the two branches, and mass-induced
sector mixing belong to the $(1+1)$-dimensional three-body construction.

\paragraph{Structural property of the internal cyclic carrier.}
The comparison of $N=1,2,3,4$ shows that $N=3$ is the smallest number
supporting distinct orientations and that one primitive three-cycle
organizes the entire minimal carrier.  The stronger statement that a
generic $N>3$ network requires additional holonomy data is a structural
observation, not a theorem excluding specially constrained single-ring
models.

\paragraph{Inherited into $(3+1)$ dimensions.}
The promotion of the two parity-paired holonomy branches to
$\mathbb C_L^3\oplus\mathbb C_R^3$ is a modeling assumption.  It is not
yet derived from a full four-dimensional microscopic theory.
\paragraph{Candidate microscopic origin of internal holonomy.}
A Lorentz-invariant six-fermion interaction constructed
from three Dirac bilinears contains two oppositely
oriented three-cycle channels.
Under suitable bilinear condensation or correlation
assumptions, these channels contribute to an effective
three-state Dirac mass structure.
This establishes an explicit operator-level realization
of cyclic three-body correlations in $(3+1)$ dimensions.
However, their matching to unitary internal transport
links and a gauge-invariant Wilson holonomy requires
additional dynamical assumptions.
The six-fermion construction should therefore be regarded
as a candidate microscopic bridge, rather than a complete
derivation of the internal gauge connection.

\paragraph{Derived inside the minimal vertical-loop effective model.}
Once the existing vertical triplets and their three transition amplitudes
are specified, Eq.~(\ref{eq:verticalcharpoly}) shows directly that the
first gauge-invariant phase dependence is proportional to the product of
three links.  The quark and lepton triplet spectra then generate the joint
holonomy potential.  No additional spectator fermion is required for
this statement.

\paragraph{Still open.}
A complete ultraviolet theory must derive the
dynamical matching between the six-fermion cyclic
channels and the effective unitary transport links,
including the emergence of a nontrivial
gauge-invariant Wilson-loop phase.

It must also establish the dynamical origin of the
internal gauge connection, derive the transition
amplitudes in Eq.~(\ref{eq:verticalMRgeneral}),
and demonstrate quantitatively that the joint
parity-symmetric potential selects one of the
asymmetric vacua in Eq.~(\ref{eq:paritypairvacua}).

Further requirements include the derivation of
the exact hypercharge embedding, anomaly
cancellation, the complete gauge-field content,
and renormalization-group consistency across
all relevant high-energy scales.

\section{Discussion}

The revised construction changes the logical status of the chiral
breaking step.  In an earlier formulation, the preferred symmetry-
breaking direction was represented effectively by a $T_8$-aligned
splitting and an adjoint vacuum.  The present formulation goes further:
$T_8$ first generates the common $2+1$ structure on both parity-related
branches, while the additional breaking of the right branch is attributed
to a holonomy effective potential generated by closed propagation of the
same vertical triplets whose third components become $D_i$ and $N_i$.
No additional spectator fermion sector is introduced.

The general vertical determinant
\begin{equation}
\det(mI-\mathcal M_V)
=ABC-A t_{23}^{2}-B t_{31}^{2}-C t_{12}^{2}
-2t_{12}t_{23}t_{31}\cos\Phi
\end{equation}
is therefore central.  It connects three-body cyclicity, the physical
vertical heavy states, and vacuum selection in a single expression.  The
phase-sensitive term contains exactly three transition amplitudes because
one primitive winding requires three links.  In the symmetric limit it
reduces to the familiar $t^3\cos\Phi$ structure.

The second important change concerns both the dimensional bridge and the
origin of left and right.  The one-dimensional configuration-space
realization cannot survive literally in $(3+1)$ dimensions because the
ordering sectors cease to be disconnected spatial sectors.  Retaining the
finite threefold holonomy spectrum therefore requires a new carrier, for
which we choose the minimal internal space $\mathbb C^3$.  The model then
does not postulate unrelated $L$ and $R$ sectors: the two opposite
holonomy phases are exchanged by parity in the massless three-body
problem, while the mass operator couples them.  Their promotion to
$\mathbb C_L^3\oplus\mathbb C_R^3$ is thus a controlled inheritance
hypothesis rather than a geometric identity.  Whether a microscopic
completion can derive this inheritance remains an open question.
A further advance is the identification of a possible
microscopic origin of the internal cyclic structure.
The Lorentz-invariant six-fermion interaction,
constructed from three Dirac bilinears, contains
two oppositely oriented three-cycle channels.
Through condensate-induced correlations, these
channels can contribute to an effective
three-state Dirac mass structure.
This provides a concrete candidate bridge between
genuine three-body interactions and the internal
holonomy assumed in the effective description.
However, the cyclic contractions of the
six-fermion operator are not themselves
unitary Wilson loops. Their identification
with internal parallel-transport links requires
additional dynamical matching, including the
generation of a nontrivial gauge-invariant
loop phase.
The present construction therefore establishes
a possible mechanism for transmitting
three-body cyclic information into the
low-energy internal sector, rather than
a complete ultraviolet derivation of the
gauge structure.

\section{Conclusion}

We have developed a framework toward a Three-Body Holonomy
Standard Model, connecting primitive three-body dynamics
to internal gauge structures, fermion mass hierarchies,
and flavor mixing.

The central observation is that a complete three-body
interaction network possesses exactly one independent
cycle, whereas smaller networks have none and larger
ones contain multiple cycles. This provides a minimal
topological motivation for selecting three as the
primitive carrier of holonomy. Higher-body interactions
may exist in the ultraviolet, but their local operators
can become increasingly irrelevant toward the infrared.

We have identified a Lorentz-invariant six-fermion
interaction as a possible microscopic source of
three-state cyclic correlations. Its matching to
unitary internal transport and Wilson holonomies
provides a candidate bridge from genuine three-body
dynamics to the internal three-state structure in
$(3+1)$ dimensions. This matching remains a hypothesis
rather than a complete ultraviolet derivation.
A hidden $SU(3)_{\rm hid}$ sector offers one possible
dynamical realization.

The two holonomy sectors $Q_3=\pm1$ of the relativistic
three-body system motivate a parity-paired internal
structure, $\mathbb C_L^3\oplus\mathbb C_R^3$.
Vertical holonomy induces a $T_8$-type $2+1$
splitting, while radiative effects and the
holonomy-dependent vacuum potential provide a
possible mechanism for selecting an asymmetric
left-right vacuum.

Horizontal holonomy organizes the three-family
mass structure. Heavy-fermion mixing generates
sector-dependent deformations, leading to physical
fermion mass hierarchies and CKM mixing.
The observable mixing arises from the misalignment
of the effective mass matrices rather than from
the common holonomy structure alone.

In the lepton sector, a restricted neutrino texture
with an additional $\mu$--$\tau$ reflection symmetry
and the phase choice $\phi=\pm2\pi/3$ yields
$\theta_{23}=45^\circ$ and
$\delta_{\rm CP}=\pm90^\circ$.
The selected benchmark gives
$\delta_{\rm CP}=270^\circ$.
This is a conditional prediction of the restricted
texture, not a unique consequence of three-body
holonomy.

The framework distinguishes the locally gauged
color and vertical sectors from the horizontal
three-state spectral space, which is not assumed
to possess an additional family gauge symmetry.
Several ingredients, particularly the ultraviolet
matching, the dynamical origin of the effective
neutrino texture, and a complete phenomenological
analysis, remain to be established.

The present results suggest that a common
three-body holonomy structure may provide an
organizing principle for gauge symmetry, family
replication, and fermion mixing.
A complete dynamical derivation and further
experimental tests are required to determine
whether this framework can be promoted to a
realistic extension of the Standard Model.

\appendix

\section{Weak-basis mass matrices and explicit transmission to CKM observables}
\label{app:weak-mass-transmission}

This appendix exhibits one traceable realization of the progression
\[
\begin{gathered}
\text{horizontal holonomy}\longrightarrow\text{heavy singlet}\longrightarrow\\
\text{effective weak-basis mass matrix}\longrightarrow\text{CKM}.
\end{gathered}
\]
The purpose is a reproducible matrix parametrization, not an additional
claim that the proposed ultraviolet theory uniquely fixes every entry.
Throughout, the horizontal index labels the three families.  The
vertical $(d,D)$ subspace is the down-type restriction of the
$(u,d,D)$ organization in Eq.~\eqref{eq:3x3VH}. It is not a second
three-family index.  We assume that $D_{L,R}$ are vector-like,
color-triplet, charge $-1/3$, $SU(2)_L$-singlet fields, so that all
blocks below can originate from gauge-invariant terms after symmetry
breaking.  In particular, the upper-right block arises from a Higgs
Yukawa interaction, while the lower-left block may be a singlet mass
mixing.  The appropriate ultraviolet representation and matching remain
to be specified.

\subsection{Stage I: common horizontal carrier and vertical parameters}

We retain the earlier horizontal operator
\begin{equation}
 C_H=\begin{pmatrix}0&1&0\\0&0&1\\e^{i\theta_H}&0&0\end{pmatrix},
 \qquad C_H^3=e^{i\theta_H}I,\qquad
 A_H=C_H+C_H^\dagger.
 \label{eq:appCH}
\end{equation}
The phase $\theta_H$ is the inherited horizontal phase in the
four-dimensional effective model. Identifying it quantitatively with
the microscopic $\theta_3$ remains a matching assumption.  
A minimal heavy-sector ansatz, consistently truncated
at the second winding, is
\begin{equation}
 M=M_0 I
 +\Delta(C_H+C_H^\dagger)
 +\Delta^{(2)}(C_H^2+C_H^{\dagger 2}),
 \label{eq:appMD}
\end{equation}
where $M_0$ denotes the heavy mass scale generated
by vertical symmetry breaking.
The coefficients $\Delta$ and $\Delta^{(2)}$
represent the first- and second-winding contributions
to the horizontal texture.


At a high energy scale, the contributions that
subsequently generate fermion masses are assumed
to share a common structure. Before the holonomy
splitting, this structure may be represented
schematically by the democratic matrix
\begin{equation}
\begin{aligned}
&
\begin{pmatrix}
u_1 & u_2 & u_3\\
d_1 & d_2 & d_3\\
D_1 & D_2 & D_3
\end{pmatrix} : \text{mass texture}
\begin{pmatrix}
s & s & s\\
s & s & s\\
s & s & s
\end{pmatrix}
\quad \text{Vertical holonomy}\longrightarrow 
\begin{pmatrix}
m & m & m\\
m & m & m\\
M & M & M
\end{pmatrix}
\end{aligned}
\label{eq:vertical_mass_splitting}
\end{equation}

The vertical holonomy thus distinguishes
the third component from the first two,
producing a $(2+1)$ structure.

Subsequently, the horizontal holonomy introduces
cyclic mixing and phase-dependent mass splittings.
In the up-type sector, the resulting mass term
takes the form
\begin{equation}
\mathcal{L}_{m}^{u}
=
-\overline{\mathbf{u}}_L
\mathcal{M}_u
\mathbf{u}_R
+\mathrm{h.c.},
\end{equation}
where
\begin{equation}
\mathbf{u}_{L,R}
=
\begin{pmatrix}
u_1\\u_2\\u_3
\end{pmatrix}_{L,R},
\end{equation}
and
\begin{equation}
\mathcal{M}_u=
\begin{pmatrix}
m &
\Delta_U+\Delta_U^{(2)}e^{-i\theta_H} &
\Delta_U e^{-i\theta_H}+\Delta_U^{(2)}
\\
\Delta_U+\Delta_U^{(2)}e^{i\theta_H} &
m &
\Delta_U+\Delta_U^{(2)}e^{-i\theta_H}
\\
\Delta_U e^{i\theta_H}+\Delta_U^{(2)} &
\Delta_U+\Delta_U^{(2)}e^{i\theta_H} &
m
\end{pmatrix}.
\label{eq:up_horizontal_mass}
\end{equation}

Here, $\Delta_U$ and $\Delta_U^{(2)}$
represent the contributions from the
first and second winding channels,
respectively, while $\theta_H$ denotes
the horizontal holonomy phase.

For real $m$, $\Delta_U$, and
$\Delta_U^{(2)}$, the matrix is Hermitian
and can be diagonalized by a unitary
transformation $F_u$:
\begin{equation}
F_u\mathcal{M}_u F_u^\dagger
=
\begin{pmatrix}
m_u & 0 & 0\\
0 & m_c & 0\\
0 & 0 & m_t
\end{pmatrix},
\label{eq:up_mass_diagonal}
\end{equation}
where the eigenvalues are taken to be
nonnegative when this single-unitary
diagonalization is used.

The corresponding mass eigenstates are
defined by
\begin{equation}
\begin{pmatrix}
u\\c\\t
\end{pmatrix}_{L,R}
=
F_u
\begin{pmatrix}
u_1\\u_2\\u_3
\end{pmatrix}_{L,R}.
\label{eq:up_mass_basis}
\end{equation}

Thus, the horizontal holonomy determines
the mass eigenstates associated with the
three up-type fermion generations.

In the down-type sector, the situation
is different because the ordinary down-type
fermions mix with heavy vector-like singlets.
The full mass matrix can be written in
block form as
\begin{equation}
\mathbb{M}_d=
\begin{pmatrix}
M_d^{(0)} & X_d\\
Y_d & M_D
\end{pmatrix},
\label{eq:down_full_matrix}
\end{equation}
where $M_d^{(0)}$ denotes the light-sector
mass matrix before heavy-state mixing,
$M_D$ is the heavy-singlet mass matrix,
and $X_d,Y_d$ describe the mixing.

For sufficiently heavy singlets,
the effective light-sector mass matrix
is approximately given by the
Schur complement,
\begin{equation}
M_d^{\mathrm{eff}}
=
M_d^{(0)}
-
X_d M_D^{-1}Y_d .
\label{eq:down_effective_mass}
\end{equation}

Consequently, the down-type mass matrix
generally differs from the up-type one,
leading to a different transformation
between the holonomy basis and the
physical mass basis.

\subsection{Stage II: explicit six-by-six weak-basis matrix}

In the common charged-current weak basis, let
\begin{equation}
 {\cal L}_{m,dD}=-\bigl(\overline d_L^{\,w},\overline D_L^{\,w}\bigr)
 \mathcal M_{dD}^{\,w}
 \begin{pmatrix}d_R^{\,w}\\D_R^{\,w}\end{pmatrix}
 +\mathrm{h.c.},\qquad
 \mathcal M_{dD}^{\,w}=\begin{pmatrix}m_d^{(0)}&X_D\\Y_D&M_D\end{pmatrix}.
 \label{eq:app6block}
\end{equation}
For an especially transparent restricted example, take
\begin{equation}
 m_d^{(0)}=\diag(a_1,a_2,a_3),\quad
 X_D=\diag(p_1,p_2,p_3),\quad
 Y_D=\diag(\mu_1,\mu_2,\mu_3),
 \label{eq:appmixdef}
\end{equation}
where $a_i=v y_{di}/\sqrt2$, $p_i=v x_i/\sqrt2$, and $\mu_i$
are mass parameters.  For the displayed texture they may be chosen
real by assumption. This is a restricted reference ansatz, not a
general rephasing theorem.  The full mass matrix reads
\begin{equation}
\mathcal M_{dD}^{\,w}=\left(\begin{array}{ccc|ccc}
 a_1&0&0&p_1&0&0\\
 0&a_2&0&0&p_2&0\\
 0&0&a_3&0&0&p_3\\ \hline
 \mu_1&0&0&M_0&A&B^*\\
 0&\mu_2&0&A^*&M_0&A\\
 0&0&\mu_3&B&A*&M_0
 \end{array}\right),
 \label{eq:app6explicit}
\end{equation}
where 
\begin{align}
  A&=\Delta_D+\Delta_D^{(2)}e^{-i\theta_H},\\
  B&=\Delta_D e^{i\theta_H} + \Delta_D^{(2)}.
\end{align}
Unlike the general ansatz $m_d^{(0)}\propto H_H$ used in the main
text, Eq.~\eqref{eq:appmixdef} chooses a particularly simple weak-basis
reference to expose the phase in $M_D$.  These are \emph{alternative
textures}, not equations that should be imposed simultaneously in an
arbitrary common weak basis.  To preserve the original horizontal
backbone literally, one may replace the upper-left diagonal block by
$m_{0d}H_H(\theta_H;r_d,s_d)$ throughout the formulas below. The
Schur-complement expressions remain unchanged.

\subsection{Stage III: eliminating the heavy singlets}

If $M_D$ is nonsingular and heavy compared with the light--heavy
mixing entries, the zero-momentum Schur complement is
\begin{equation}
 M_d^{\rm eff}=m_d^{(0)}-X_DM_D^{-1}Y_D,
 \label{eq:appschur}
\end{equation}
up to higher-order normalization and momentum-dependent corrections.
Writing $R=M_D^{-1}$ gives the directly inspectable $3\times3$ matrix
\begin{equation}
M_d^{\rm eff}=\begin{pmatrix}
 a_1-p_1\mu_1R_{11}&-p_1\mu_2R_{12}&-p_1\mu_3R_{13}\\
 -p_2\mu_1R_{21}&a_2-p_2\mu_2R_{22}&-p_2\mu_3R_{23}\\
 -p_3\mu_1R_{31}&-p_3\mu_2R_{32}&a_3-p_3\mu_3R_{33}
 \end{pmatrix}.
 \label{eq:app3explicit}
\end{equation}
Let $F_H$ denote the unitary eigenvector matrix of $C_H$, with
$F_H^\dagger C_HF_H=\diag(e^{i\varphi_0},e^{i\varphi_1},e^{i\varphi_2})$
and $\varphi_k=(\theta_H+2\pi k)/3$.  Then

\begin{equation}
 R=F_H
 \operatorname{diag}
 \left(
 \frac{1}{\Lambda_0},
 \frac{1}{\Lambda_1},
 \frac{1}{\Lambda_2}
 \right)F_H^\dagger,
\end{equation}
where
\begin{equation}
 \Lambda_k=M_0
 +2\Delta_D\cos\varphi_k
 +2\Delta_D^{(2)}\cos2\varphi_k,
 \qquad
 \varphi_k=\frac{\theta_H+2\pi k}{3}.
 \label{eq:appresolvent}
\end{equation}
so that the phase dependence of each matrix element is explicit:
\begin{equation}
 R_{ij}=\sum_{k=0}^2\frac{(F_H)_{ik}(F_H)^*_{jk}}{\Lambda_k}.
 \label{eq:appRij}
\end{equation}

For
$|\Delta_D|,|\Delta_D^{(2)}|\ll |M_0|$,
the inverse heavy mass matrix admits
the following expansion is
\begin{equation}
\begin{aligned}
 M_D^{-1}={}&\frac{I}{M_0}
 -\frac{\Delta_D}{M_0^2}
 (C_H+C_H^\dagger)\\
 &-\frac{\Delta_D^{(2)}}{M_0^2}
 (C_H^2+C_H^{\dagger2})
 +O\left(\frac{\Delta_{\rm max}^2}{M_0^3}\right),
\end{aligned}
\label{eq:appRexp}
\end{equation}
where
$\Delta_{\rm max}=
\max(|\Delta_D|,|\Delta_D^{(2)}|)$.
This implies, for example,

\begin{equation}
\begin{aligned}
(M_d^{\rm eff})_{12}
&\simeq
\frac{p_1\mu_2}{M_0^2}
\left(\Delta_D+
\Delta_D^{(2)}e^{-i\theta_H}\right),\\
(M_d^{\rm eff})_{23}
&\simeq
\frac{p_2\mu_3}{M_0^2}
\left(\Delta_D+
\Delta_D^{(2)}e^{-i\theta_H}\right),\\
(M_d^{\rm eff})_{31}
&\simeq
\frac{p_3\mu_1}{M_0^2}
\left(\Delta_De^{i\theta_H}
+\Delta_D^{(2)}\right).
\end{aligned}
\label{eq:appphasetransmission}
\end{equation}
These phases may be redistributed among links by a diagonal family
rephasing, but the oriented product of the three links retains their
total loop phase.  A nonzero phase in the down matrix alone does not
establish a physical CKM phase. The relative up/down left rotations
must be tested.

\subsection{Stage IV: mass-basis rotations and CKM transmission}

Starting from $M_u^{\rm eff}$ and Eq.~\eqref{eq:app3explicit}, form
\begin{equation}
 {\cal H}_q=M_q^{\rm eff}M_q^{{\rm eff}\dagger},\qquad
 U_{qL}^\dagger{\cal H}_q U_{qL}=\diag(m_{q1}^2,m_{q2}^2,m_{q3}^2),
 \quad q=u,d.
 \label{eq:appH}
\end{equation}
For the Hermitian up-type mass matrix,
the left-handed diagonalization convention
is related to the transformation introduced
in Stage I by
\begin{equation}
U_{uL}=F_u^\dagger,
\end{equation}
up to eigenstate ordering and unphysical
phase conventions. Therefore,
\begin{equation}
V_{\mathrm{CKM}}^{(0)}=F_uU_{dL}.
\end{equation}
In the absence of light--heavy wave-function corrections,
$V^{(0)}=U_{uL}^\dagger U_{dL}$.  In the common horizontal eigenbasis,
$\widetilde{\cal H}_q=F_H^\dagger{\cal H}_qF_H$, first-order
nondegenerate perturbation theory gives
\begin{equation}
 (\Omega_q)_{ij}\simeq
 \frac{(F_H^\dagger\delta{\cal H}_qF_H)_{ij}}
 {m_{qj}^{(0)2}-m_{qi}^{(0)2}},\quad i\ne j,
 \qquad V^{(0)}\simeq I+\Omega_d-\Omega_u.
 \label{eq:appCKMlin}
\end{equation}
This perturbative expression applies when both zeroth-order matrices
share $F_H$ and eigenvalue gaps are nonzero.  For the alternative
diagonal example of Eq.~\eqref{eq:appmixdef}, use its own zeroth-order
basis instead. One should not mix these two perturbative conventions.
The observable CP test is
\begin{equation}
 J_{\rm CKM}=Im(V_{us}V_{cb}V_{ub}^*V_{cs}^*),
 \qquad Im\Tr([\mathcal H_u,\mathcal H_d]^3)\ne0
 \label{eq:appJ}
\end{equation}
for CP violation with nondegenerate quark masses.  Fitting the six
masses and four independent CKM observables is necessary before
identifying a predictive $\theta_H$ range.

Using the diagonalization conventions introduced
above, the unitary CKM matrix is
$V_{\mathrm{CKM}}^{(0)}=F_uU_{dL}$.
We now display its individual entries
and the corrections induced by heavy-singlet mixing.

To display the individual CKM entries,
consider small relative rotations in a
common reference basis.
Writing
\begin{equation}
V_{\mathrm{CKM}}^{(0)}
\simeq I+\omega,
\qquad
\omega^\dagger=-\omega,
\end{equation}
with
\begin{equation}
\omega=\Omega_d-\Omega_u,
\end{equation}
one obtains
\begin{equation}
V_{\mathrm{CKM}}^{(0)}
\simeq
\begin{pmatrix}
1 & \omega_{12} & \omega_{13}\\
-\omega_{12}^* & 1 & \omega_{23}\\
-\omega_{13}^* & -\omega_{23}^* & 1
\end{pmatrix}.
\label{eq:app_ckm_entries}
\end{equation}

The independent off-diagonal entries are
\begin{align}
V_{us}^{(0)}&\simeq
(\Omega_d)_{12}-(\Omega_u)_{12},
\\
V_{cb}^{(0)}&\simeq
(\Omega_d)_{23}-(\Omega_u)_{23},
\\
V_{ub}^{(0)}&\simeq
(\Omega_d)_{13}-(\Omega_u)_{13}.
\end{align}

The heavy down-type singlets introduce
an additional correction through their
mixing with the light fermions.
Defining
\begin{equation}
\Theta_L\simeq X_D M_D^{-1},
\end{equation}
the effective charged-current mixing
matrix is
\begin{equation}
V_{\mathrm{light}}
\simeq
F_u
\left(
I-\frac12\Theta_L\Theta_L^\dagger
\right)
U_{dL}.
\label{eq:app_ckm_heavy}
\end{equation}
Introducing the Hermitian correction matrix
\begin{equation}
\eta=
\frac12 F_u\Theta_L\Theta_L^\dagger
F_u^\dagger,
\qquad
\eta^\dagger=\eta,
\end{equation}
we obtain
\begin{equation}
V_{\mathrm{light}}
\simeq
(I-\eta)V_{\mathrm{CKM}}^{(0)}.
\label{eq:app_ckm_eta}
\end{equation}
Neglecting products of $\eta$ with the
small relative family rotations, the
individual entries become
\begin{equation}
V_{\mathrm{light}}
\simeq
\begin{pmatrix}
1-\eta_{11}
&
\omega_{12}-\eta_{12}
&
\omega_{13}-\eta_{13}
\\
-\omega_{12}^*-\eta_{21}
&
1-\eta_{22}
&
\omega_{23}-\eta_{23}
\\
-\omega_{13}^*-\eta_{31}
&
-\omega_{23}^*-\eta_{32}
&
1-\eta_{33}
\end{pmatrix}.
\label{eq:app_ckm_components}
\end{equation}
In particular,
\begin{align}
V_{us}&\simeq\omega_{12}-\eta_{12},
\\
V_{cb}&\simeq\omega_{23}-\eta_{23},
\\
V_{ub}&\simeq\omega_{13}-\eta_{13},
\\
V_{tb}&\simeq1-\eta_{33}.
\end{align}

The quantities $\omega_{ij}$ describe
the relative rotations induced by the
up- and down-type mass structures,
whereas $\eta_{ij}$ describes the
non-unitary corrections associated
with heavy-singlet mixing.
These expressions are valid at leading
order in the relative rotations and
heavy-light mixing. Quadratic family
rotation effects must be retained
for a precision analysis, particularly
for the diagonal CKM entries and
the small element $V_{ub}$.

\subsection{Stage V: heavy-state non-unitarity and limitations}

The heavy-singlet corrections derived in
Stage IV modify the unitarity of the
effective light-sector charged-current matrix.

At the same order, the light kinetic operators
and the definition of the effective mass matrix
require consistent canonical normalization.
The full six-state singular-value decomposition
is the appropriate benchmark when the hierarchy
is marginal.
Non-unitarity, induced $Z$-mediated
flavor-changing neutral currents, and the
$Zb\bar b$ constraint must be checked against data.
The restricted ansatz above has not been subjected
here to a simultaneous numerical quark-mass,
CKM, and precision-electroweak fit;
it illustrates transmission, not an established
fit or a derived CP prediction.

\section{Acknowledgments}

During the preparation of this manuscript, as a non-native English speaker,
the author used ChatGPT (OpenAI GPT-6 ) for language editing, improving clarity 
and organization, and assisting with the presentation of the manuscript. 
All scientific arguments, calculations, interpretations, and conclusions 
were reviewed and verified by the author, who takes full responsibility 
for the content of the manuscript.

All numerical inputs and results supporting the findings of this study 
are provided in the article and its appendices. No additional datasets were generated.


\begin{thebibliography}{99}
\bibitem{Munakata1990}
Y.~Munakata, J.~Sakamoto, T.~Ino, F.~Nagamura, and F.~Yamamoto,
``Exact Solutions of a Relativistic Three-Body Bound State in $1+1$ Dimensions,''
Prog. Theor. Phys. \textbf{83}, 84--96 (1990),
doi:10.1143/PTP.83.84.

\bibitem{Sakamoto1993}
J.~Sakamoto,
Construction of $N$ body bound state solution of the Dirac particles in $1+1$ dimensional space-time,
Prog. Theor. Phys. 89 (1993) 119--130.

\bibitem{SakamotoNakanoYoshikawa1993}
J.~Sakamoto, T.~Nakano, and T.~Yoshikawa,
``Two Types of N-Body Bound State Solutions of the Dirac Particles with the $\delta$ Function Potential in 1+1 Dimensional Space-Time,''
Prog. Theor. Phys. \textbf{89}, 767--770 (1993),
doi:10.1143/PTP.89.767.

\bibitem{Ohya2024}
S.~Ohya,
``Topologically nontrivial three-body contact interaction in one dimension,''
Prog. Theor. Exp. Phys. \textbf{2024}, 013A01 (2024),
doi:10.1093/ptep/ptad149, arXiv:2310.16576.

\bibitem{YoshikawaPhase2026}
T.~Yoshikawa,
``Phase Encoding of Genuine Three-Body Interactions in a Relativistic
Dirac System in $(1+1)$ Dimensions,''
arXiv:2609.25478 (2026).

\bibitem{YoshikawaFamily2026}
T.~Yoshikawa,
``Three-Body Holonomy as a Toy-Model Mechanism for Family Triplication
and Mass Hierarchy in $(1+1)$ Dimensions,''
arXiv:2609.24022 (2026).

\bibitem{YoshikawaSU32026}
T.~Yoshikawa,
``Three-Body Holonomy and $SU(3)\to SU(2)\times U(1)$ Symmetry Breaking,''
arXiv:2610.00235 (2026).

\bibitem{YoshikawaComposite2026}
T.~Yoshikawa,
``A Composite Three-Body-Holonomy Toy Framework for Dynamical
Electroweak-Scale Selection,''
arXiv:2610.04202 (2026).

\bibitem{KobayashiMaskawa1973}
M.~Kobayashi and T.~Maskawa,
``CP-Violation in the Renormalizable Theory
of Weak Interaction,''
Prog. Theor. Phys. \textbf{49}, 652--657 (1973),
doi:10.1143/PTP.49.652.

\bibitem{MNS1962}
Z.~Maki, M.~Nakagawa and S.~Sakata,
``Remarks on the Unified Model of Elementary Particles,''
Prog. Theor. Phys. \textbf{28}, 870--880 (1962),
doi:10.1143/PTP.28.870.


\bibitem{Simon1983}
B.~Simon,
``Holonomy, the quantum adiabatic theorem, and Berry's phase,''
Phys. Rev. Lett. \textbf{51}, 2167--2170 (1983),
doi:10.1103/PhysRevLett.51.2167.

\bibitem{Berry1984}
M.~V.~Berry,
``Quantal phase factors accompanying adiabatic changes,''
Proc. R. Soc. Lond. A \textbf{392}, 45--57 (1984),
doi:10.1098/rspa.1984.0023.

\bibitem{WilczekZee1984}
F.~Wilczek and A.~Zee,
``Appearance of Gauge Structure in Simple Dynamical Systems,''
Phys. Rev. Lett. \textbf{52}, 2111--2114 (1984),
doi:10.1103/PhysRevLett.52.2111.

\bibitem{Hosotani1983}
Y.~Hosotani,
``Dynamical mass generation by compact extra dimensions,''
Phys. Lett. B \textbf{126}, 309--313 (1983),
doi:10.1016/0370-2693(83)90170-3.

\bibitem{Hosotani1984}
Y.~Hosotani,
``Dynamical gauge-symmetry breaking and left-right asymmetry in
higher-dimensional theories,''
Phys. Rev. D \textbf{29}, 731--745 (1984),
doi:10.1103/PhysRevD.29.731.

\bibitem{Lin1997}
F.~J.~Lin,
``Symplectic reduction, geometric phase, and internal dynamics in
three-body molecular dynamics,''
Phys. Lett. A \textbf{234}, 291--300 (1997),
doi:10.1016/S0375-9601(97)00608-7.

\bibitem{Bedaque1999}
P.~F.~Bedaque, H.-W.~Hammer, and U.~van Kolck,
``Renormalization of the Three-Body System with Short-Range Interactions,''
Phys. Rev. Lett. \textbf{82}, 463--467 (1999),
doi:10.1103/PhysRevLett.82.463, arXiv:nucl-th/9809025.

\bibitem{BraatenHammer2006}
E.~Braaten and H.-W.~Hammer,
``Universality in few-body systems with large scattering length,''
Phys. Rept. \textbf{428}, 259--390 (2006),
doi:10.1016/j.physrep.2006.03.001, arXiv:cond-mat/0410417.

\bibitem{NuFIT2025}
I.~Esteban, M.~C.~Gonzalez-Garcia, M.~Maltoni, I.~Martinez-Soler,
J.~P.~Pinheiro, and T.~Schwetz,
``NuFit-6.0: Updated global analysis of three-flavor neutrino oscillations,''
JHEP \textbf{12}, 216 (2024), arXiv:2410.05380,
together with the NuFIT 6.1 (2025) parameter update at
\url{https://www.nu-fit.org/}.

\bibitem{Branco2024}
J.~M.~Alves, G.~C.~Branco, A.~L.~Cherchiglia, C.~C.~Nishi,
J.~T.~Penedo, P.~M.~F.~Pereira, M.~N.~Rebelo, and J.~I.~Silva-Marcos,
``Vector-like singlet quarks: A roadmap,''
Phys. Rept. \textbf{1057}, 1--69 (2024),
doi:10.1016/j.physrep.2023.12.004, arXiv:2304.10561.

\bibitem{Wilson1974}
K.~G.~Wilson,
``Confinement of quarks,''
Phys. Rev. D \textbf{10}, 2445--2459 (1974),
doi:10.1103/PhysRevD.10.2445.

\bibitem{tHooft1976}
G.~'t Hooft,
``Computation of the quantum effects due to a
four-dimensional pseudoparticle,''
Phys. Rev. D \textbf{14}, 3432--3450 (1976),
doi:10.1103/PhysRevD.14.3432.

\bibitem{Klevansky1992}
S.~P.~Klevansky,
``The Nambu--Jona-Lasinio model of quantum
chromodynamics,''
Rev. Mod. Phys. \textbf{64}, 649--708 (1992),
doi:10.1103/RevModPhys.64.649.

\bibitem{AntuschHinzeSaad2026}
S.~Antusch, K.~Hinze, and S.~Saad,
``Updated running quark and lepton parameters at various scales,''
Phys. Rev. D \textbf{113}, 095011 (2026),
doi:10.1103/fdcc-ycph.

\bibitem{Jarlskog1985}
C.~Jarlskog,
``Commutator of the Quark Mass Matrices in the
Standard Electroweak Model and a Measure of
Maximal CP Nonconservation,''
Phys. Rev. Lett. \textbf{55}, 1039--1042 (1985),
doi:10.1103/PhysRevLett.55.1039.

\end{thebibliography}
\end{document}